\documentclass[aps, prb, twocolumn,showkeys, showpacs,amsmath,amssymb,
superscriptaddress]{revtex4-1}
\usepackage{graphicx}
\usepackage{dcolumn}
\usepackage{bm}
\usepackage{epsfig}

\begin{document}
\title{Klein Tunneling in the presence of random impurities}

\author{S. Palpacelli} \email{silviapalpacelli@gmail.com}
\affiliation{Numidia s.r.l.,
  Via Giacomo Peroni, 130, 00131, Roma, Italy.}

\author{M. Mendoza} \email{mmendoza@ethz.ch} \affiliation{ ETH
  Z\"urich, Computational Physics for Engineering Materials, Institute
  for Building Materials, Schafmattstrasse 6, HIF, CH-8093 Z\"urich,
  Switzerland.}

\author{H. J. Herrmann}\email{hjherrmann@ethz.ch} \affiliation{ ETH
  Z\"urich, Computational Physics for Engineering Materials, Institute
  for Building Materials, Schafmattstrasse 6, HIF, CH-8093 Z\"urich,
  Switzerland.}

\author{S. Succi} \email{succi@iac.cnr.it} \affiliation{Istituto per
  le Applicazioni del Calcolo C.N.R., Via dei Taurini, 19 00185, Rome
  Italy,\\and Freiburg Institute for Advanced Studies, Albertstrasse,
  19, D-79104, Freiburg, Germany.}

\date{\today}
\begin{abstract}
  In this paper, we study Klein tunneling in random media.  
  To this purpose, we simulate the propagation of a relativistic Gaussian
  wavepacket through a graphene sample with randomly distributed
  potential barriers (impurities).  The simulations, based on a
  relativistic quantum lattice Boltzmann method, permit to compute the
  transmission coefficient across the sample, thereby providing an
  estimate for the conductivity as a function of impurity
  concentration and strength of the potentials. It is found that the
  conductivity loss due to impurities is significantly higher for
  wave-packets of massive particles, as compared to massless ones. 
  A general expression for the loss of conductivity as a function of the
  impurity percentage is presented and successfully compared with the
  Kozeny-Carman law for disordered media in classical fluid dynamics.
\end{abstract}



\maketitle

\section{Introduction}

As opposed to classical quantum mechanics where electrons tunneling
into a barrier are exponentially damped, relativistic scattering was
shown by Klein in 1929 \cite{KLEIN} to follow a very unexpected
behavior: If the potential is of the order of the electron mass or
higher the barrier becomes virtually transparent to the
electrons. This is called the Klein paradox. Experimental realizations
were not available until the recent discovery of graphene
\cite{Natletter,Geim1}. This material has revealed a series of amazing
properties, such as ultra-high electrical conductivity, ultra-low
shear viscosity to entropy ratio, as well as exceptional structural
strength, as combined with mechanical flexibility and optical
transparency. Many of these fascinating properties are due to the fact
that, consisting of literally one single carbon monolayer, graphene
represents the first instance ever of a truly two-dimensional material
(the ``ultimate flatland'' \cite{PhysToday}). Moreover, due to the
special symmetries of the honeycomb lattice, electrons in graphene are
shown to behave like an effective Dirac fluid of {\it massless} chiral
quasi-particles, propagating at a Fermi speed of about $v_F \sim c/300
\sim 10^6$ m/s. This configures graphene as an unique,
slow-relativistic electronic fluid, where many unexpected
quantum-electrodynamic phenomena can take place, \cite{QGP}. For
instance, since electrons are about $300$ times slower than photons,
their mutual interaction is proportionately enhanced, leading to an
effective fine-structure constant $\alpha_{gr} = e^2/\hbar v_F \sim
1$.  As a result of such strong interactions, it has been recently
proposed that this peculiar 2D graphene electron gas should be
characterized by an exceptionally low viscosity/entropy ratio
(near-perfect fluid), coming close to the famous AdS-CFT lower bound
conjectured for quantum-chromodynamic fluids, such as quark-gluon
plasmas \cite{QGP}.  This spawns the exciting prospect of observing
electronic pre-turbulence in graphene samples, as first pointed out in
Ref. \cite{MULLER} and confirmed by recent numerical simulations
\cite{MENDOZA}.

The zero-mass of electronic excitations in graphene may have other
spectacular consequences.  For instance, it has been recently pointed
out \cite{NATURE} that graphene could offer the first experimental of
the so-called Klein paradox, i.e. the capability of quantum
wavefunctions to undergo zero reflection from a potential barrier much
higher than the energy of the wavefunction itself. This property,
which relies exclusively upon the spinorial nature of the Dirac
wavefunction, stands in stark contrast with the corresponding
non-relativistic behavior, which predicts an exponential decay of the
transmission coefficient with the difference $V_0-E$, $V_0$ being the
height of the barrier and $E$ the wavefunction energy.  Based on an
analytical solution of the scattering problem for a monochromatic
plane wave, the authors were able to show that, depending on a series
of geometrical and energy parameters, special angles of incidence
(resonant angles) provide literally zero reflectivity: the plane wave
goes completely across the barrier.  Besides its intellectual charm,
such property is of great practical interest for the study of
electronic transport in graphene \cite{PRB1,PRB2}, and it is expected
to play an important role in the understanding of the minimum
conductivity of graphene \cite{RMP2011}. Furthermore, the electronic
spectrum of graphene can change depending on the substrate, for
instance on SiC the energy spectrum presents a gap of width $2mv_F^2$,
which makes it possible to model the electric transport by using the
massive Dirac equation \cite{massive, massive2}. Therefore, it is also
interesting to study the Klein tunneling in a random media for this
kind of gaped-samples (massive fermions case).

On the other hand, due to the fact that, under suitable conditions
\cite{MULLER}, electronic excitations in graphene behave as an
effective relativistic Dirac fluid, in the presence of a random media,
transport laws similar to the ones ruling fluid motion in diluted
porous media, may be expected to apply. We refer here, e.g. to the
Carman-Kozeny law \cite{rumer, Bear}, which relates the permeability
of a porous medium (conductivity of a graphene sample) to the solid
concentration (impurity density).

The paper is organized as follows: first, we introduce a brief
description of the quantum lattice Boltzmann (QLB) method \cite{QLB};
second, we study the case of Klein tunneling of a Gaussian wave packet
through a rectangular potential barrier. Subsequently, we present
numerical solutions of the Dirac equation in the presence of random
impurities, thereby providing an estimate for the effects of the
impurity concentration on the conductivity of the graphene sample, for
both cases, massless and massive Dirac fermions. The simulations are
performed using a QLB model, which is also introduced as a new tool to
study transport phenomena in graphene. Finally, we discuss and
summarize the results.

\section{The quantum lattice Boltzmann method}
\begin{figure}
\centering
\includegraphics[scale=0.45]{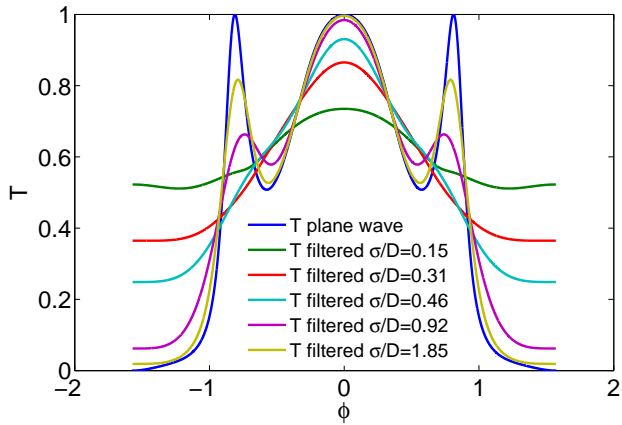}
\caption{The transmission coefficient of a Gaussian wavepacket, as
  computed with the analytical convolution, Eq. \eqref{eq:TGAUSS}, as a
  function of the incidence angle $\phi$ for
  $\sigma/D=0.15,0.31,0.46,0.92,1.85$. The blue line corresponds to
  the unfiltered case, $\sigma \rightarrow \infty$, corresponding to a
  plane wave.  }
\label{fig:Fig2}
\end{figure}
\begin{figure}
\centering
\includegraphics[scale=0.4]{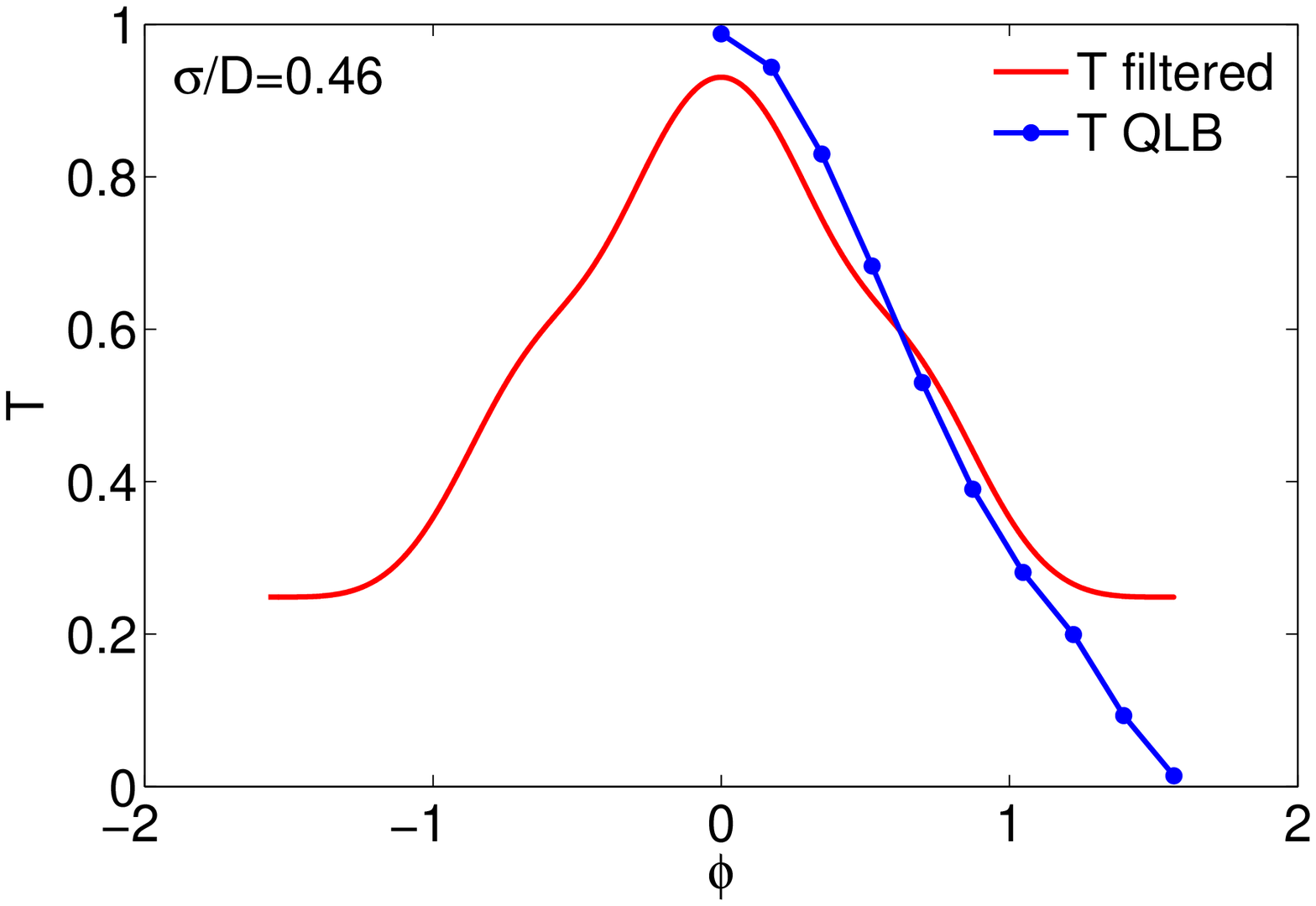}
\includegraphics[scale=0.4]{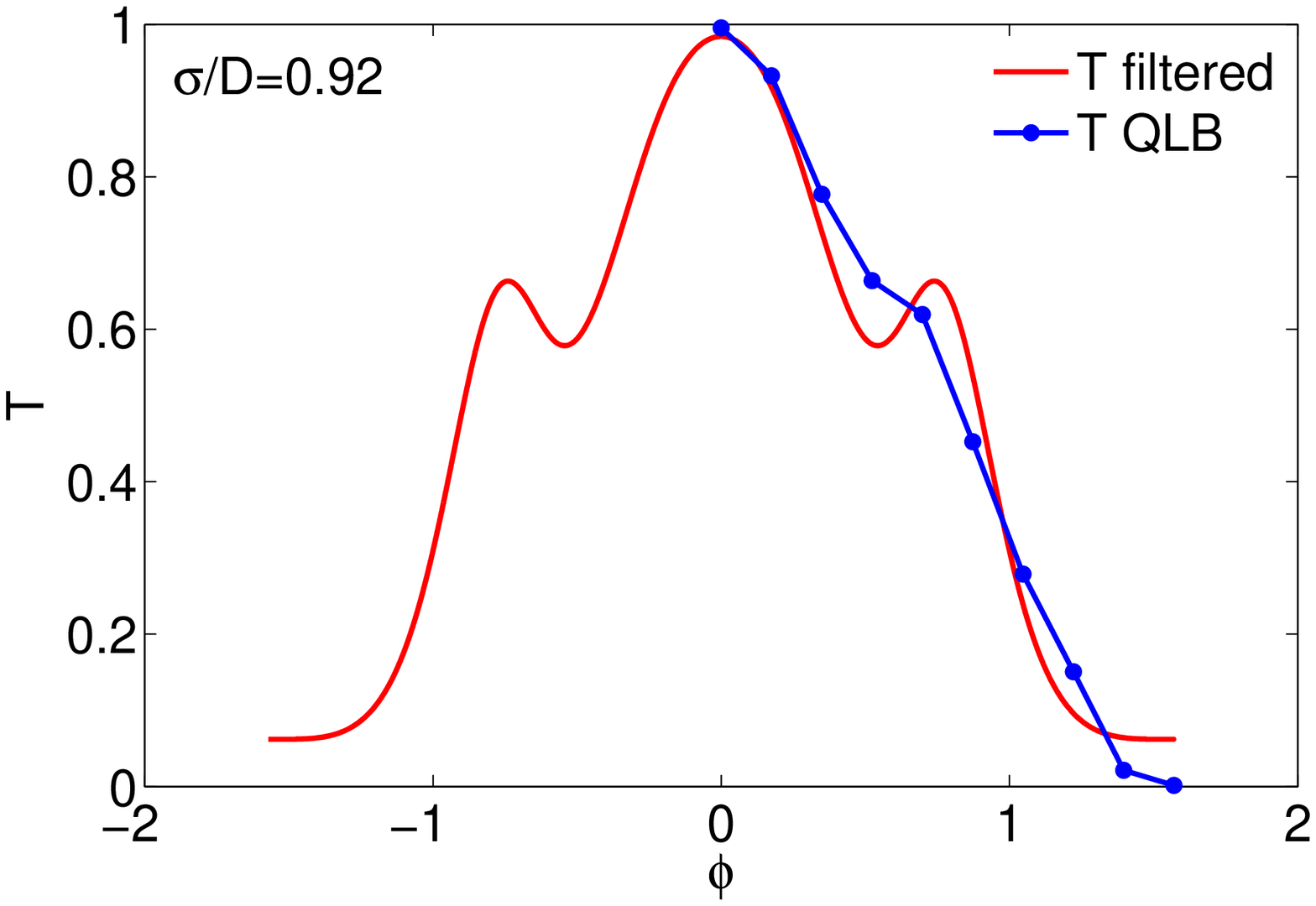}\\
\includegraphics[scale=0.4]{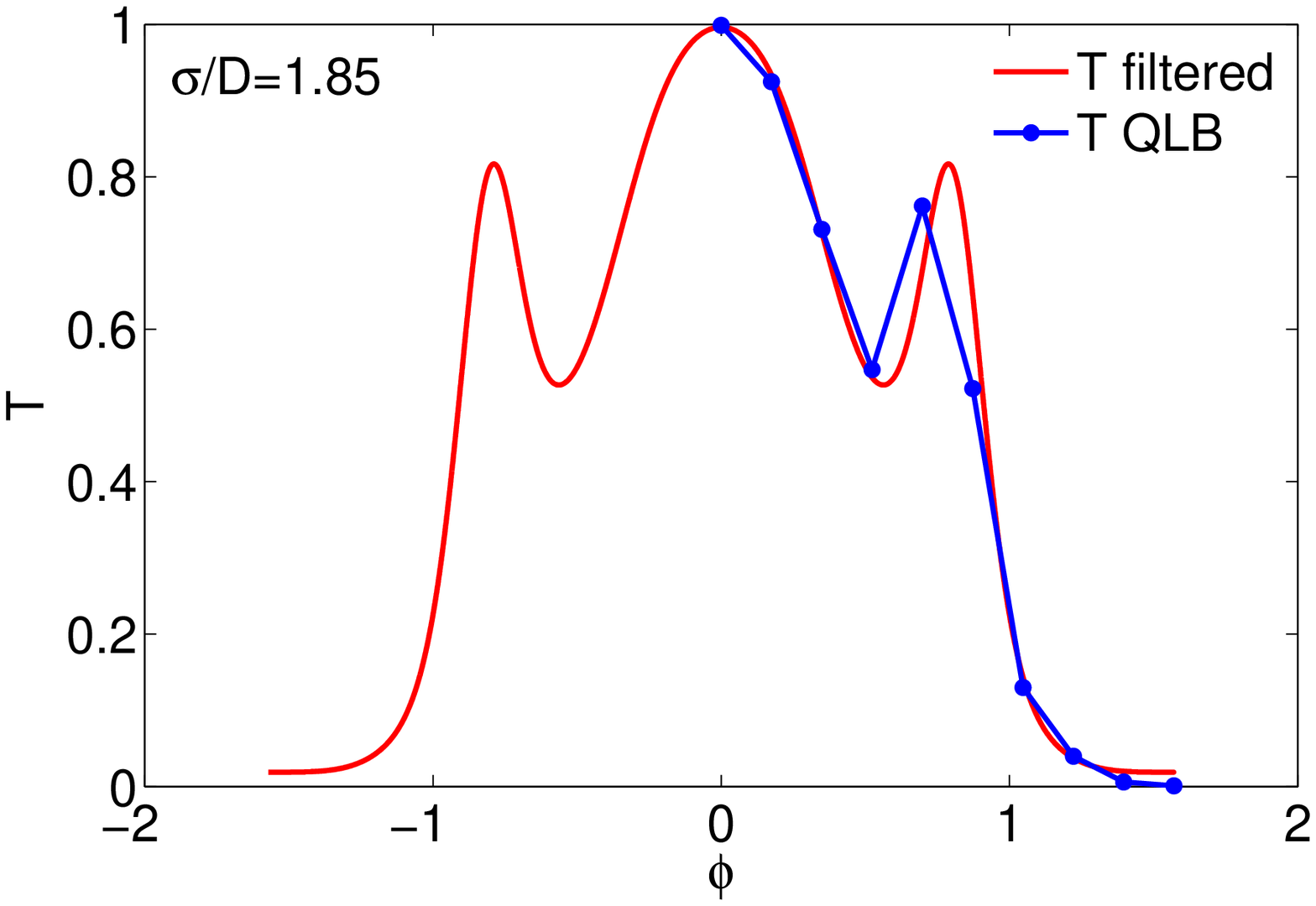}
\caption{The transmission coefficient of a Gaussian wavepacket as a
  function of the incidence angle $\phi$ for $\sigma=24$, $48$ and
  $96$ (in lattice units), corresponding to $\sigma/D=0.46,0.92,1.85$,
  as computed via convolution (solid line) and by QLB simulations
  (line with dots).  }
\label{fig:Fig3}
\end{figure}
\begin{figure}
\centering
\includegraphics[width=0.15\textwidth]{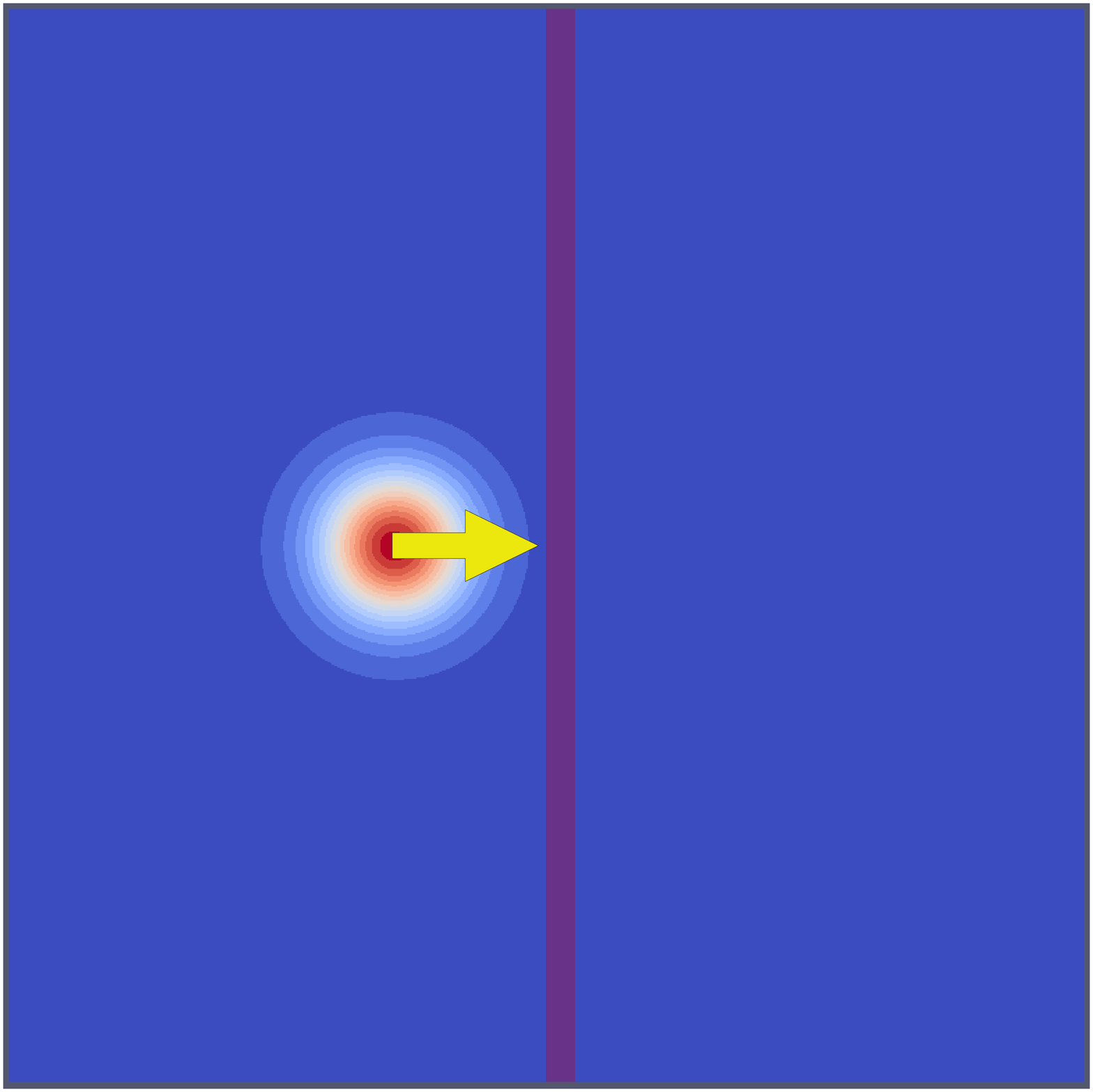}\hspace{0.1cm}
\includegraphics[width=0.15\textwidth]{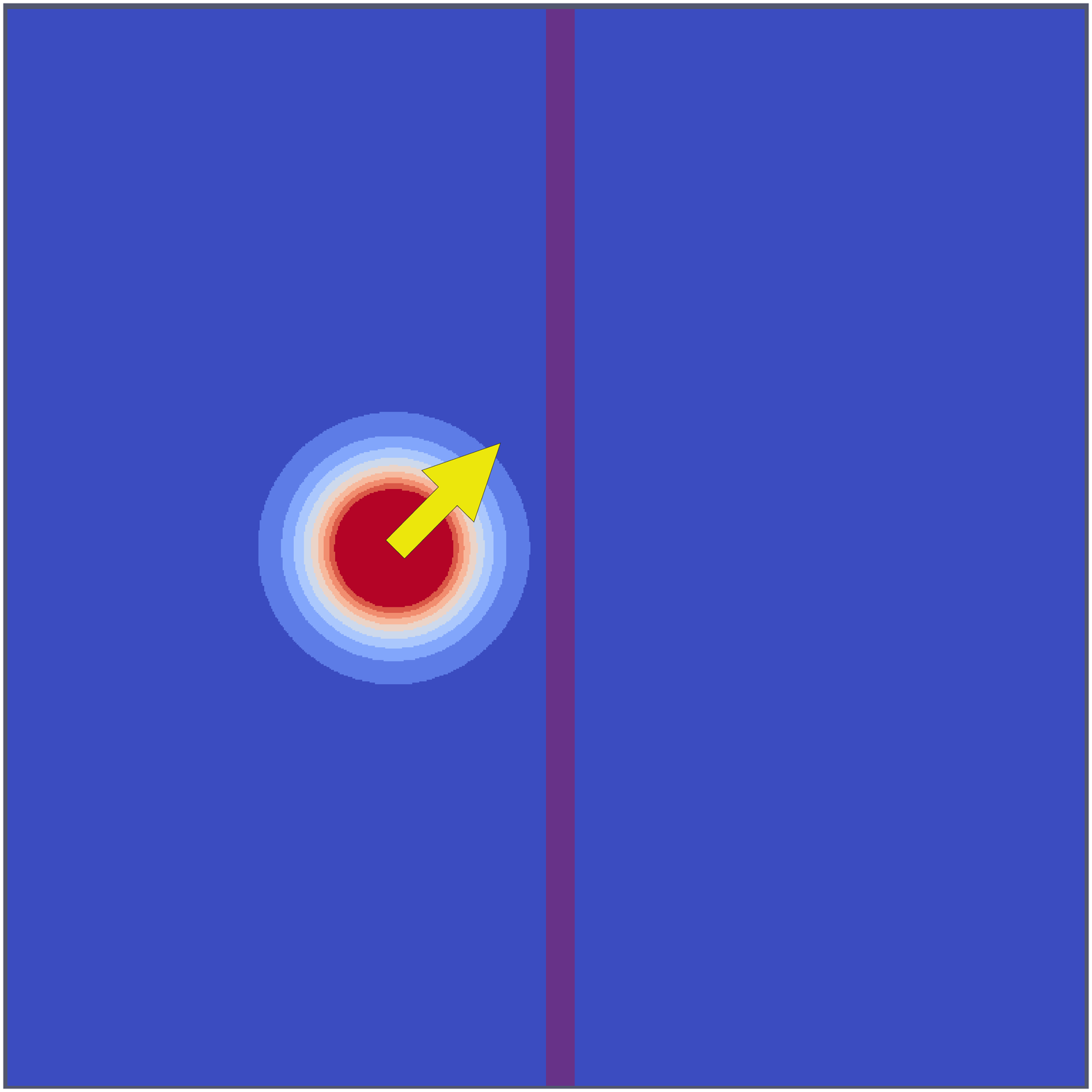}\hspace{0.1cm}
\includegraphics[width=0.15\textwidth]{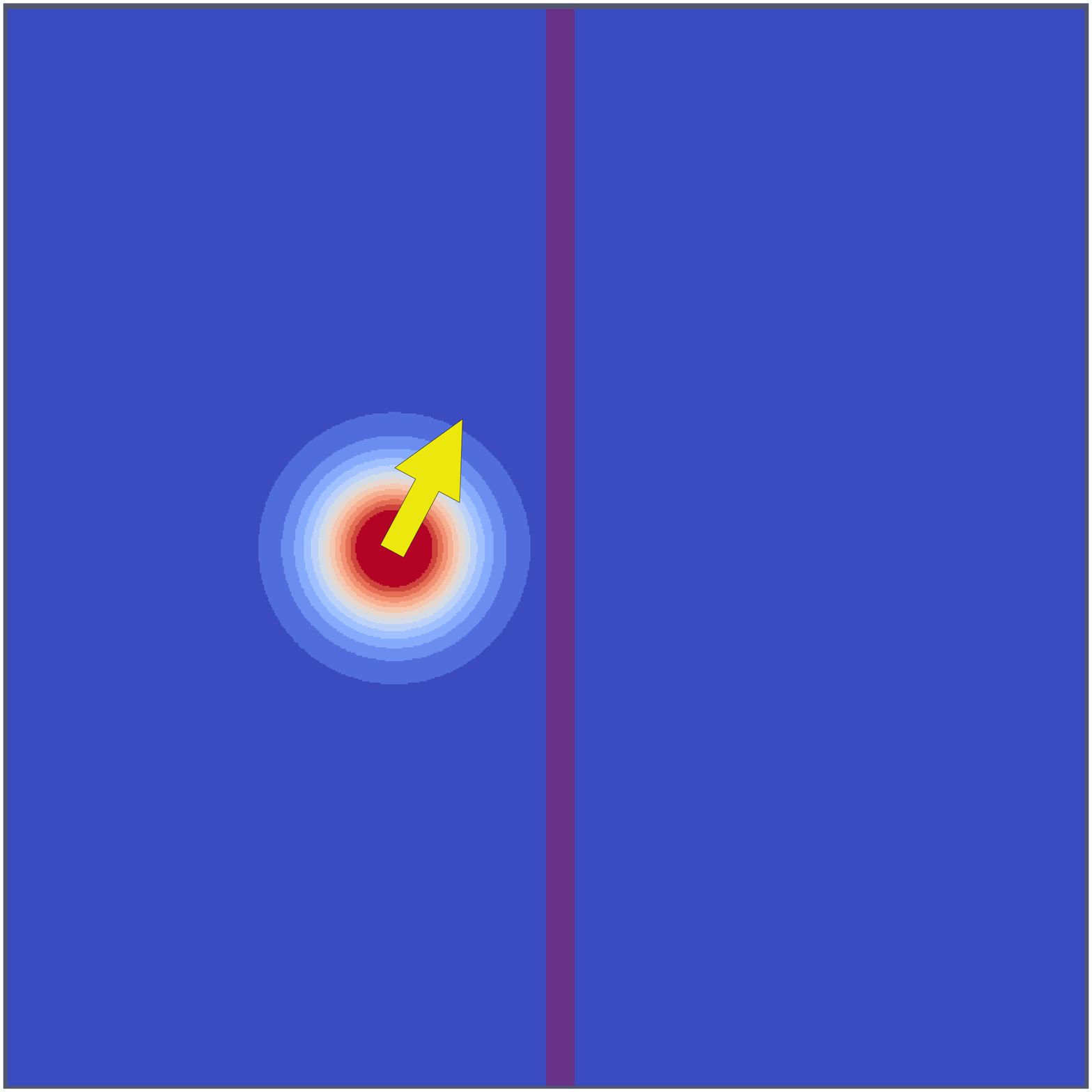}\\\vspace{0.1cm}
\includegraphics[width=0.15\textwidth]{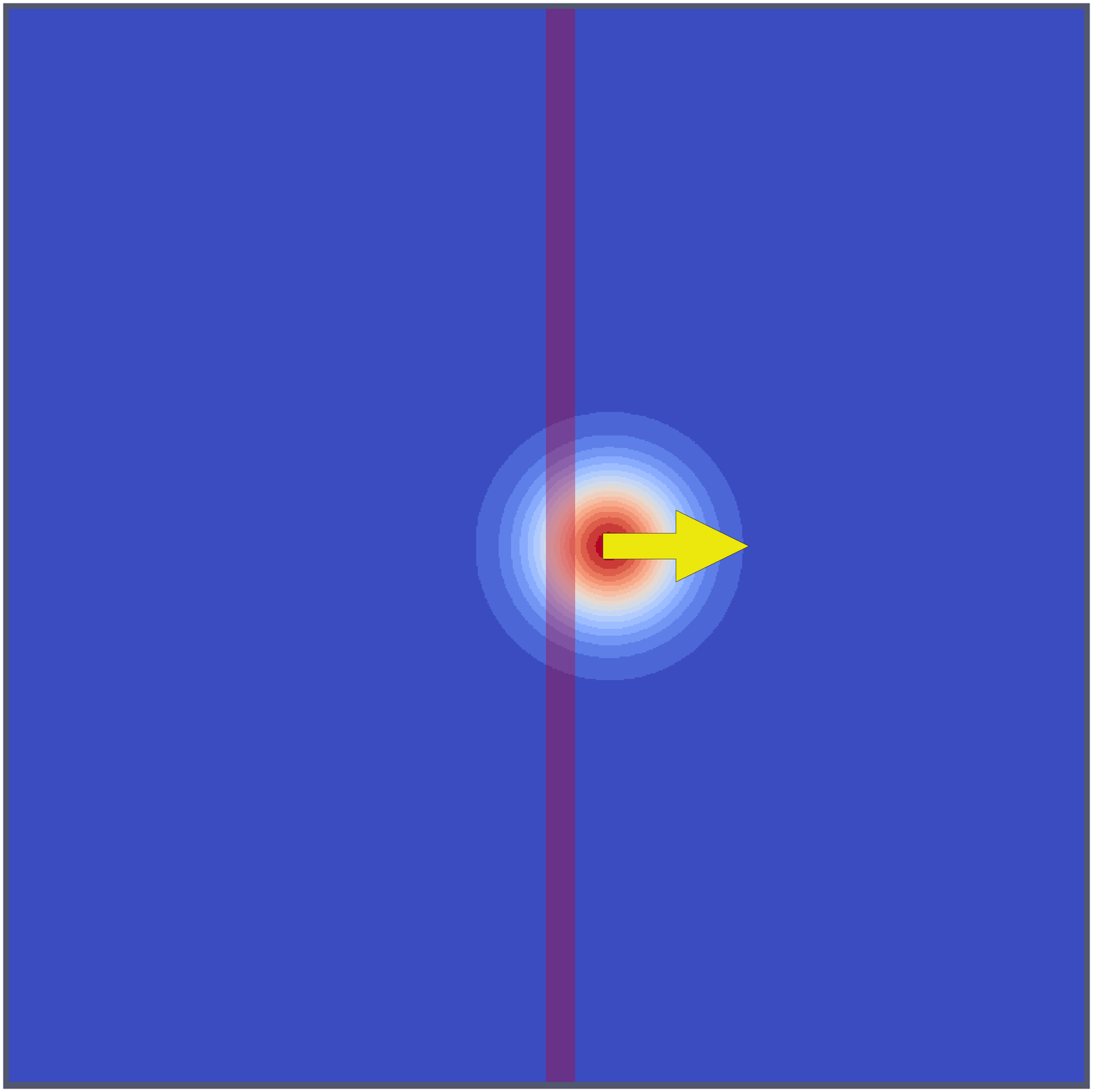}\hspace{0.1cm}
\includegraphics[width=0.15\textwidth]{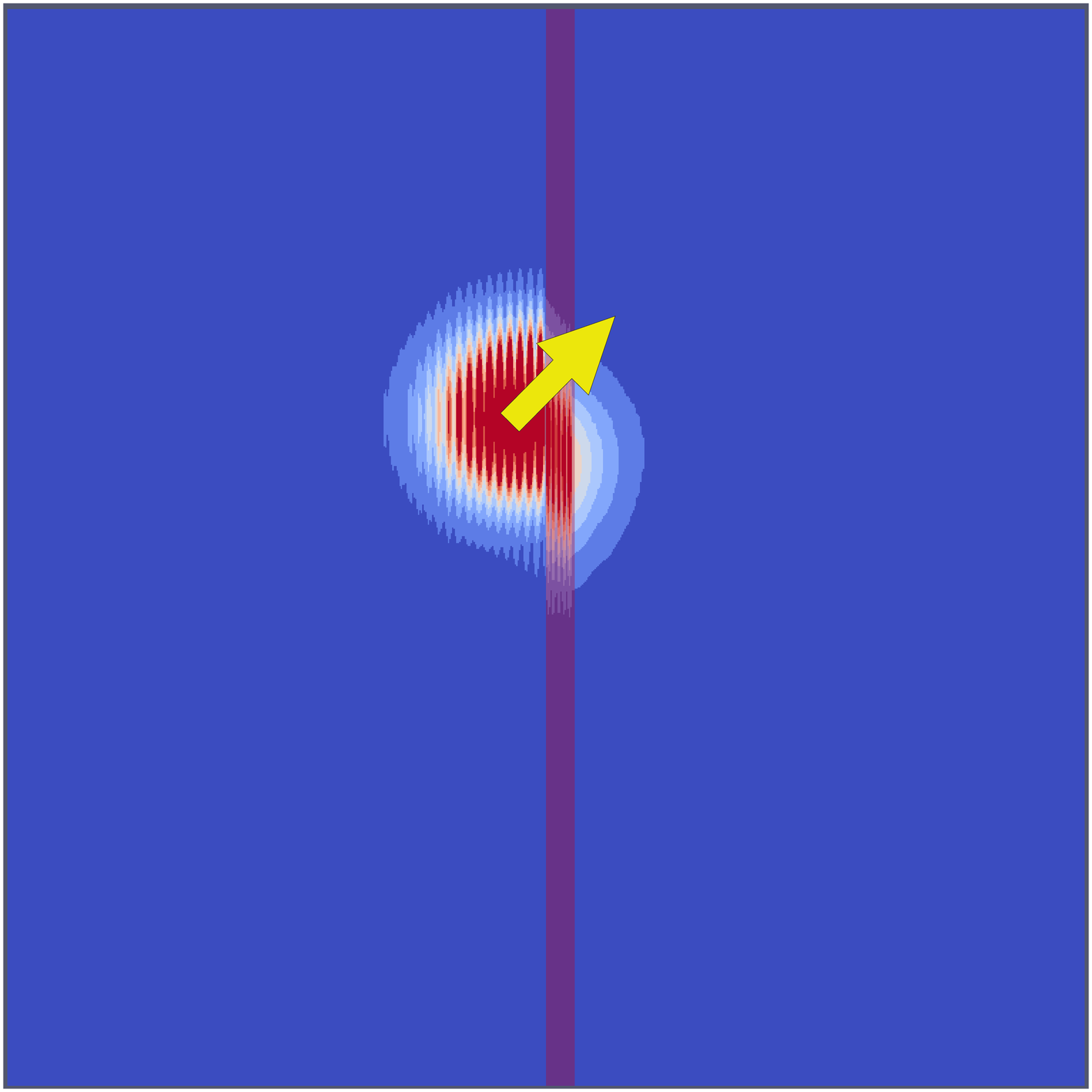}\hspace{0.1cm}
\includegraphics[width=0.15\textwidth]{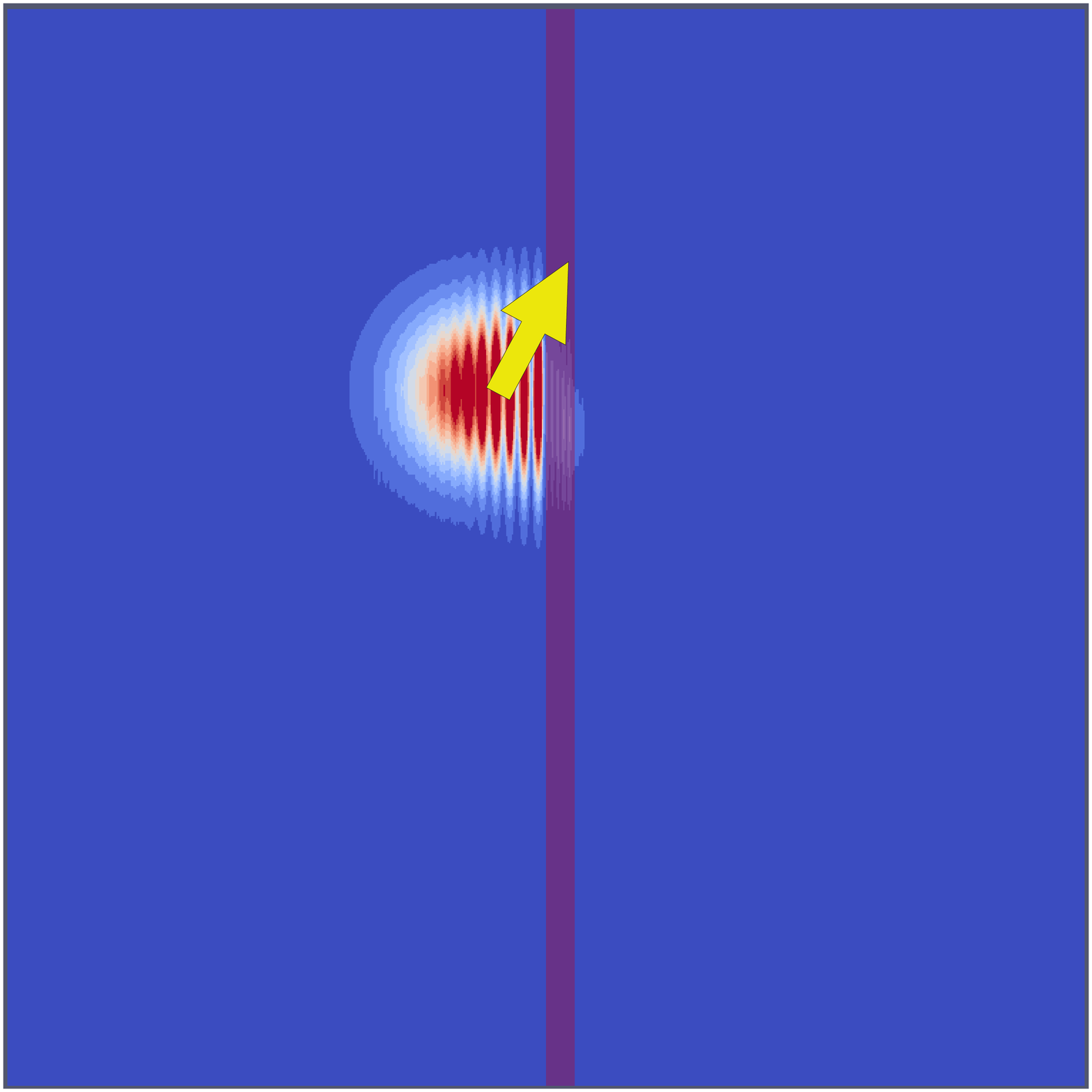}\\\vspace{0.1cm}
\includegraphics[width=0.15\textwidth]{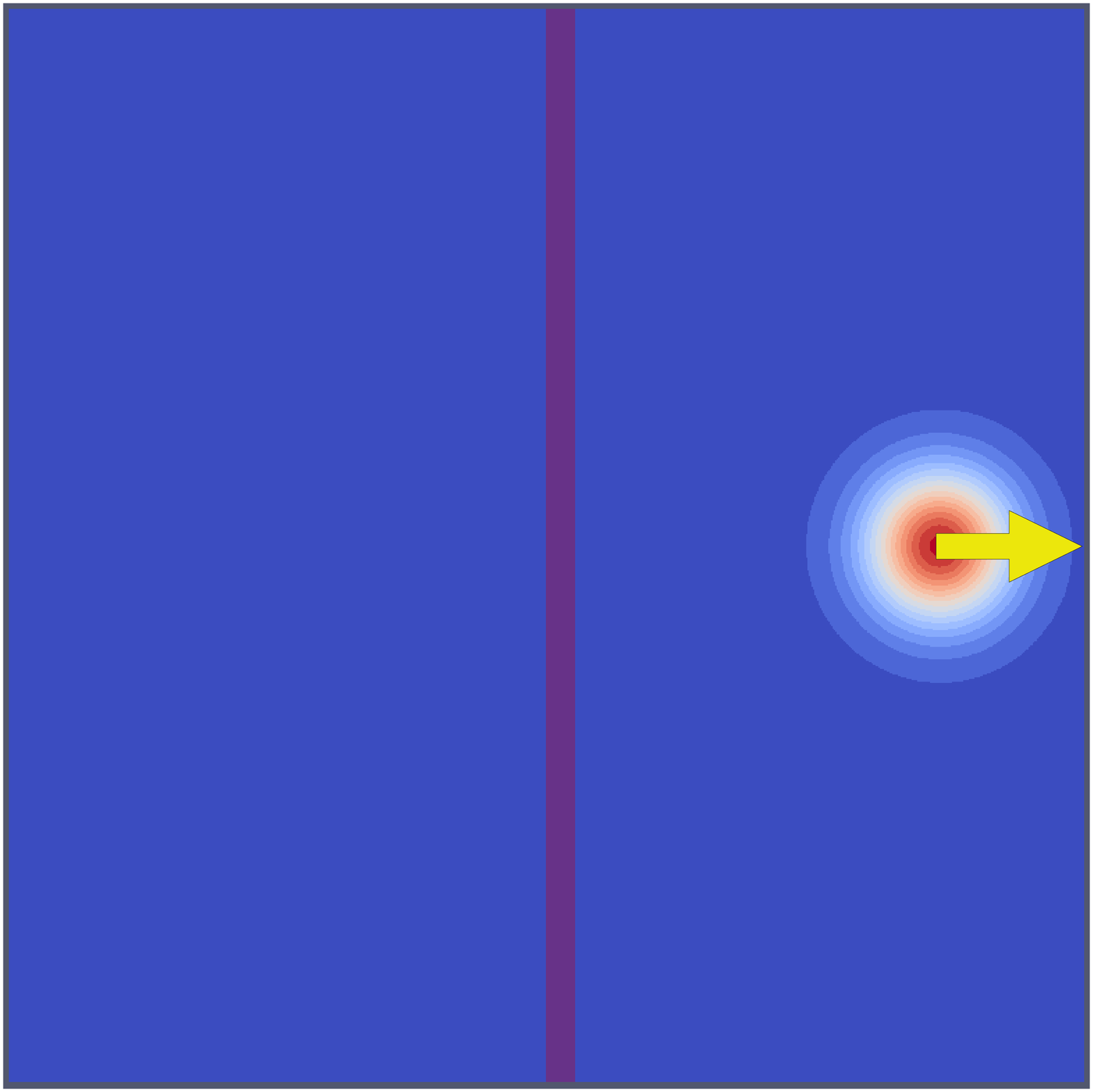}\hspace{0.1cm}
\includegraphics[width=0.15\textwidth]{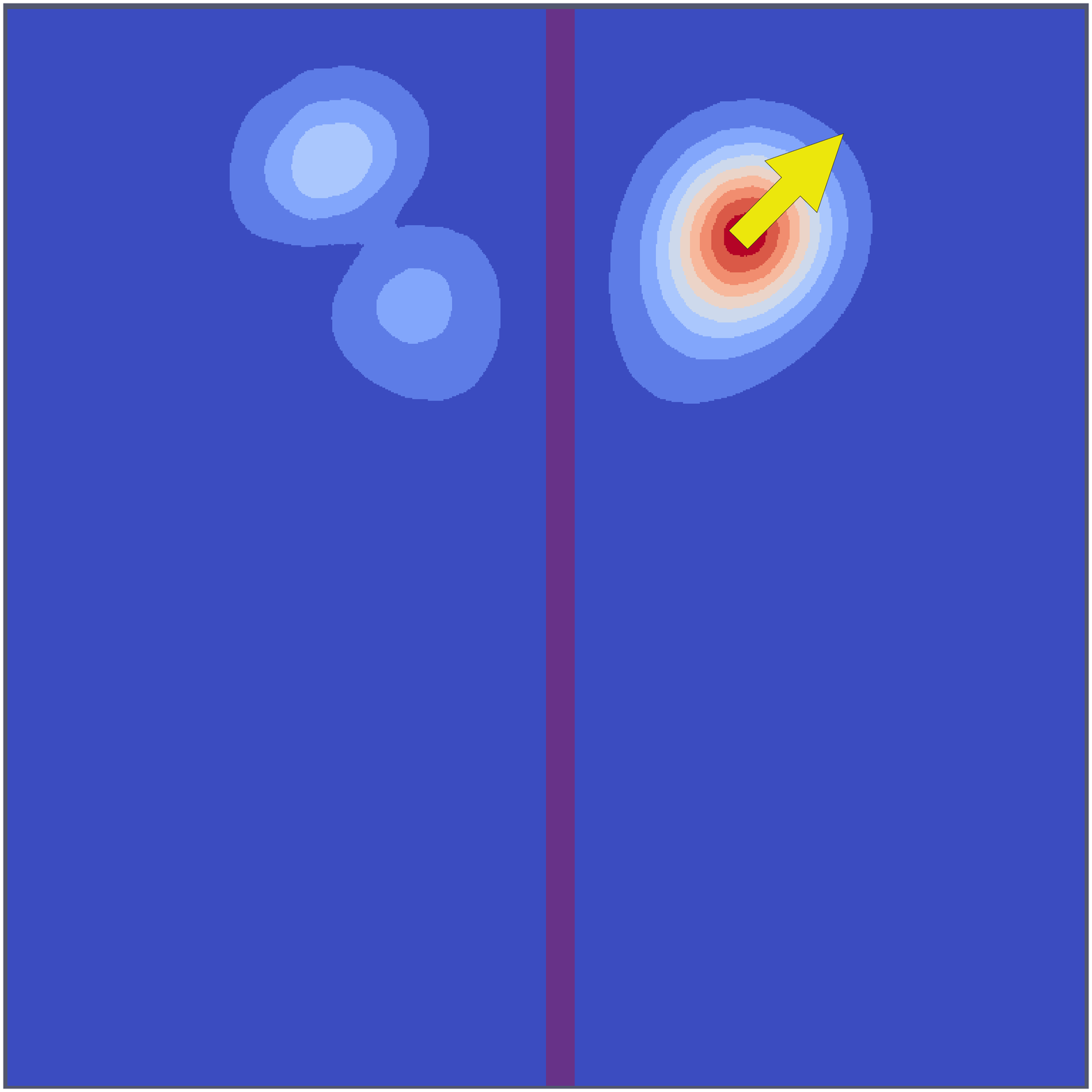}\hspace{0.1cm}
\includegraphics[width=0.15\textwidth]{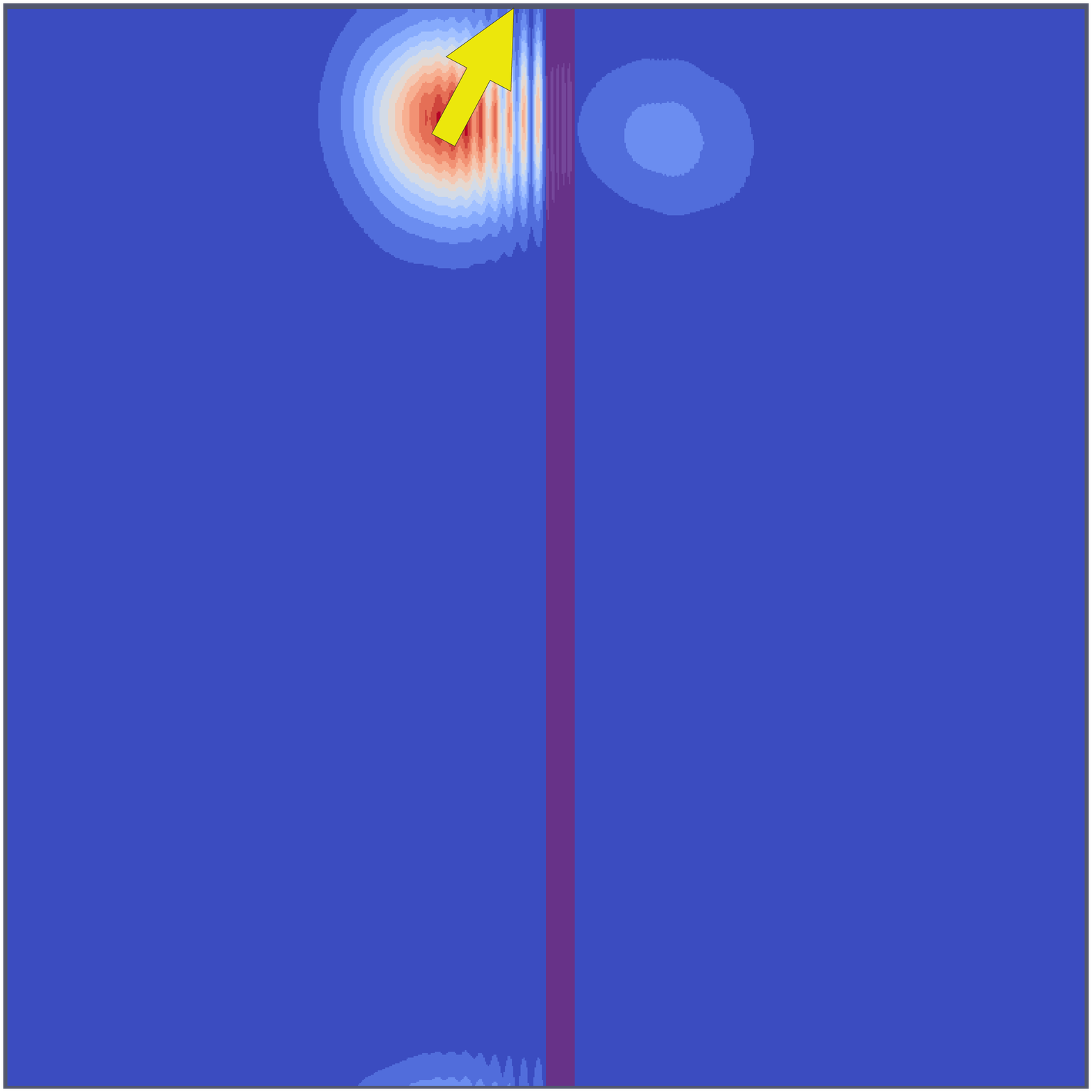} 
\caption{Snapshots of the wavepacket density at various instants,
  $t=0, 420, 1050$ (lattice units), for the case $\phi=0$ (left) and
  $\phi=2 \pi/9$ (middle), and $\phi=\pi/3$ (right) for
  $\sigma/D=1.85$.  In the middle, as one can see, after significant
  distortion in the intermediate stage of the evolution, the
  wavepacket manages to be transmitted across the barrier to a
  substantial extent ($T=0.76$). On the other hand, at the right, the
  packet is mostly bounced-back by the barrier, with transmission
  coefficient as low as $T = 0.13$. For visualization purposes, the
  color bar scale has been modified independently for each figure.}
\label{fig:Fig5}
\end{figure}

The quantum lattice Boltzmann (QLB) method \cite{QLB} is a
quantum-kinetic technique that was originally devised for
non-relativistic quantum problems and recently shown to provide a
second-order accurate solver for relativistic wave scattering and
propagation \cite{DELLAR}. Since the method is relatively new in the
relativistic context, for the sake of self-containedness, we revisit
here its main technical aspects.  For full details, see our recent
work \cite{DL,PS08}.

The quantum lattice Boltzmann equation was initially derived from a
formal parallel between the kinetic lattice Boltzmann equation and the
single-particle Dirac equation \cite{BSV,QLB,NOTE}.  For our purpose,
it proves expedient to transform the standard form of the Dirac
equation into the Majorana form, in which all matrices are real
\cite{BerestetskiiPitaevskiiLifshitz82},
\begin{equation} 
\left[ \partial_t  + c (-\alpha^x \partial_x + \beta \partial_y -
\alpha^z \partial_z )  + i \omega_c \alpha^y  - i g I \right] \psi = 0,
\label{eq:Majorana}
\end{equation}
This form is obtained by multiplying the standard Dirac equation on
the left and right by the involution matrix $U = 2^{-1/2} (\alpha^y +
\beta)$.  In the above, $c$ is the light speed, $\hbar$ is the reduced
Planck's constant, $I$ is the identity operator, and $\omega_c =
mc^2/\hbar$ is the Compton frequency for a particle of mass $m$.  The
wavefunction $\psi$ is a complex four-spinor, and $\alpha$ and $\beta$
are the standard Dirac matrices.  The last term couples the
wavefunction to an applied scalar potential $V(x,y,z)$ via the
coefficient $g=q V/\hbar$, where $q$ is the electric charge
\cite{BerestetskiiPitaevskiiLifshitz82}. Note that since the spin
states mix-up during propagation (spinning particles), there is no
basis in which all three matrices $\alpha^x$, $\alpha^y$, $\alpha^z$
are simultaneously diagonal.

Let us consider a one-dimensional version of the Dirac equation.  In
particular, let $Z$ be a unitary matrix, diagonalizing the streaming
matrix $-\alpha^z$:
\begin{equation}
Z = \frac{1}{\sqrt{2}}\begin{pmatrix}
  0 & -1 & 0 & 1\\
  1 & 0 & -1 & 0\\
  0 & 1 & 0 & 1\\
  1 & 0 & 1 & 0
    \end{pmatrix}.
\label{eq:Zdef}
\end{equation}
Applying the matrix $Z$ to Eq. \eqref{eq:Majorana}, the streaming matrix
along $z$ is diagonalized and the collision matrix is also transformed
accordingly
\begin{equation}\label{eq:streamZ}
  \begin{aligned}
    \big[ \partial_t &+ c Z^{-1} (-\alpha^z) Z \partial_z \\ &+
    Z^{-1}(- c \alpha^x \partial_x + c \beta \partial_y + i \omega_c
    \alpha^y - i g I) Z \big] Z^{-1} \psi = 0.
  \end{aligned}
\end{equation}
Neglecting any dependence of $\psi$ on the $x$ and $y$ coordinates,
Eq. \eqref{eq:streamZ} may be written as a pair of one-dimensional Dirac
equations
\begin{equation} \label{eq:dirac-1D}
\begin{split}
\partial_t u_{1,2} + c \partial_z u_{1,2} & = \omega_c d_{2,1} + i g  u_{1,2},\\
\partial_t d_{1,2} - c \partial_z d_{1,2} & = - \omega_c u_{2,1} + i g  d_{1,2},
\end{split}
\end{equation}
for the variables $(u_1,d_2)$ and $(u_2,d_1)$ that represent the
rotated wavefunction $Z^{-1} \psi = (u_1,u_2,d_1,d_2)^T$. The
components $u$ and $d$ propagate up and down the $z$ axis
respectively, and the subscripts indicate the spin up (1) and spin
down (2) states, respectively.
The system of Eq. \eqref{eq:dirac-1D} may be treated as a Boltzmann
equation for a pair of complex distribution functions $u_{1,2}$ and
$d_{1,2}$\cite{QLB}.  Equation \eqref{eq:dirac-1D} may thus be
discretized using the same approach as in lattice Boltzmann method,
i.e. by integrating along the characteristic light-cones $dz = \pm c
dt$.

The resulting system of algebraic equations reads as follows
\begin{equation} \label{eq:qLB-1D}
\begin{split}
\widehat{u}_{1,2} - u_{1,2} & = \frac{1}{2} \widetilde{m} (d_{2,1} +
\widehat{d}_{2,1}) + \frac{1}{2} i \widetilde{g} (u_{1,2} + \widehat{u}_{1,2}), \\
\widehat{d}_{1,2} - d_{1,2} & = - \frac{1}{2} \widetilde{m} (u_{2,1} +
\widehat{u}_{2,1}) + \frac{1}{2} i \widetilde{g} (d_{1,2} + \widehat{d}_{1,2}),
\end{split}
\end{equation}
where the hat superscript ($\,\,\widehat{}\,\,$) indicates that the
wavefunction is evaluated at the end-point of the corresponding
streaming step, namely
\begin{equation} \label{eq:qLB-notation}
\begin{split}
\widehat{u}_{1,2} &= u_{1,2}(z + \Delta z, t + \Delta t), \quad u_{1,2} = u_{1,2}(z,t)\\
\widehat{d}_{1,2} &= d_{1,2} (z - \Delta z, t + \Delta t), \quad d_{1,2} = d_{1,2}(z,t).
\end{split}
\end{equation}
The dimensionless Compton frequency is $\widetilde{m} = \omega_c
\Delta t$, and the dimensionless scalar potential is $\widetilde{g} =
g(z,t) \Delta t$.

The pair of equations \eqref{eq:qLB-1D} can be solved algebraically,
delivering explicit expressions for $\widehat{u}_{1,2}$ and
$\widehat{d}_{1,2}$:
\begin{equation} \label{eq:exp-qLB1D}
\begin{split}
\widehat{u}_{1,2} &= a u_{1,2} + b d_{2,1},\\
\widehat{d}_{1,2} &= a d_{1,2} - b u_{2,1},
\end{split}
\end{equation}
where the coefficients $a$ and $b$ are
\begin{equation*}
a = \frac{1 - \Omega/4}{1 + \Omega/4 -  i \widetilde{g}} , \quad b =
\frac{\widetilde{m}}{1 + \Omega/4 -  i \widetilde{g}}, \quad \Omega = \widetilde{m}^2 - \widetilde{g}^2.
\end{equation*}
These coefficients satisfy $|a|^2+|b|^2=1$, so that the right hand
side of Eq. \eqref{eq:exp-qLB1D} corresponds to multiplying the rotated
wavefunction $Z^{-1} \psi=(u_1,u_2,d_1,d_2)^T$ by the unitary
collision matrix
\begin{equation} \label{eq:coll_mat}
Q = \begin{pmatrix}
  a & 0 & 0 & b\\
  0 & a & b & 0\\
  0 & -b & a & 0\\
  -b & 0 & 0 & a
\end{pmatrix}.
\end{equation}
The streaming step propagates $u_{1,2}$ upwards and $d_{1,2}$
downwards, along the light cones given by $\Delta z = \pm c \Delta t$.
Note that this unitary operation is numerically {\it exact}, without
round-off error, because the distribution function is integrally
transferred from the source to the destination site, and no fractional
transport is involved.  Since both streaming and collisions step are
unitary, the overall QLB scheme evolves the discrete wavefunction
through a sequence of unitary operations for any value of the discrete
time step $\Delta t$. In addition, since streaming proceeds upwind
only (no centered spatial differences) along the discrete light-cones
associated with each component $\Psi_i$ , the QLB dispersion relation
is automatically free from fermion-doubling, \cite{TworzPRB08}.  This,
together with the excellent efficiency of the method, especially on
parallel computers \cite{PARALB}, should make QLB a potentially
appealing candidate for computational studies of electron transport in
graphene.

The scheme extends to multiple dimensions through an operator
splitting technique.  Within this method, the three-dimensional Dirac
equation splits into the sum of three one-dimensional equations, each
involving spatial derivatives along one single direction.  Each of the
three stages representing evolution by a timestep $dt$ is accomplished
by rotating $\psi$ to diagonalise the relevant streaming matrix,
taking one timestep of the existing one-dimensional QLB scheme
described above, and rotating $\psi$ back to its original basis.  The
algorithm is thus composed of the following three steps: 1) Rotate
$\psi$ with $X^{-1}$, collide with $X^{-1} \widehat{Q} X$, stream
along $x$, rotate back with $X$; 2) Rotate $\psi$ with $Y^{-1}$,
collide with $Y^{-1} \widehat{Q} Y$, stream along $y$, rotate back
with $Y$; 3) Rotate $\psi$ with $Z^{-1}$, collide with $Z^{-1}
\widehat{Q} Z$, stream along $z$, rotate back with $Z$.  This form
emphasizes the symmetry between the three steps, but since the
streaming matrix along $y$ is already diagonal in the Majorana form,
$Y=I$ is the identity matrix.  The matrix $X$ reads as follows:
\begin{equation}
X = \frac{1}{\sqrt{2}} \begin{pmatrix}
-1 & 0 & 1 & 0 \\
0 & 1 & 0 & -1 \\
1 & 0 & 1 & 0 \\
0 & 1 & 0 & 1 
\end{pmatrix},
\end{equation}
and the $Z$ matrix is given in Eq. \eqref{eq:Zdef} above.

The collision term splits into three parts, each of which is combined
with the corresponding streaming step.  The collision matrix thus
coincides, up to a unitary transformation, with the collision matrix
for the one-dimensional QLB scheme, with a timestep $\frac{1}{3} dt$
(see Ref. \cite{PS08}). In particular, $\widehat{Q}$ is given by
\begin{equation} \label{eq:coll_mat3D}
\widehat{Q} = \begin{pmatrix}
  \widehat{a} & 0 & 0 & -\widehat{b}\\
  0 & \widehat{a} & \widehat{b} & 0\\
  0 & -\widehat{b} & \widehat{a} & 0\\
  \widehat{b} & 0 & 0 & \widehat{a}
\end{pmatrix},
\end{equation}
where the coefficients
\begin{equation*}
\widehat{a} = \frac{1-\Omega_3/4}{1 + \Omega_3/4 - i\widetilde{g}_3}, 
\quad \widehat{b} = \frac{\widetilde{m}_3}{1 + \Omega_3/4 - i\widetilde{g}_3},
\end{equation*}
are written in terms of the rescaled dimensionless Compton and
potential frequencies
\begin{equation*}
\Omega_3 = \widetilde{m}^2_3 - \widetilde{g}_3^2, \quad 
\widetilde{m}_3 = \frac{1}{3} \omega_c dt, \quad \widetilde{g}_3 =\frac{1}{3} g dt.
\end{equation*} 
The pattern of $+$ and $-$ signs in the $\widehat{b}$ terms on the
off-diagonal of $\widehat{Q}$ follows the same pattern as the
$\alpha^y$ matrix. The rotated matrices $X^{-1} \widehat{Q} X$ and
$Z^{-1} \widehat{Q} Z$ have the same sign pattern as $Q$, but
$\widehat{Q}$ does not.

Summarizing, QLB provides a unitary, explicit algorithm for quantum
wavefunctions in which information propagates along classical
trajectories represented by a sequence of three one-dimensional
light-cones, thereby avoiding any mixing of the spinorial components
during the streaming step. Although detailed comparisons with other
techniques remain to be developed, there are reasons to believe that
such simplification may result in enhanced computational efficiency,
especially with parallel computers in mind.  Finally, we wish to point out
that the same algorithm describes both relativistic and
non-relativistic quantum wavepackets, depending on the value of the
mass $m$ and the characteristic strength of the potential energy.

\section{Relativistic Gaussian wavepackets}
\begin{figure}
   \centering   
   \includegraphics[scale=0.38]{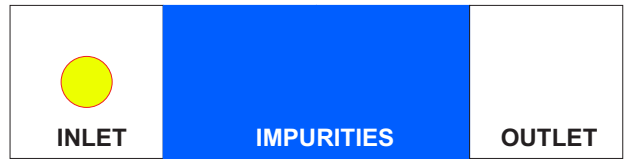}    
   \caption{Sketch of the domain setting used in our simulations of
     the propagation of a Gaussian wave packet through a porous
     medium.}
   \label{fig:domain}
\end{figure}

Since we are interested in applications on graphene, hereafter, our
simulations will be performed in two spatial dimensions,  
(for more details see Ref.~\cite{PS08}).
The propagation of a plane wave through a rectangular potential
barrier was discussed in Ref.~\cite{NATURE}. However due to the fact
that it only applies to monochromatic plane waves, i.e. infinitely
extended states which may not necessarily be realized under all experimental conditions , it 
is of interest to explore to what extent are such results affected by the finite extent of the wavefunction. 
Here, for simplicity, we consider a Gaussian wavepacket of the form
\begin{equation}
\psi_l(x,y) = \frac{A_k}{{(4\pi \sigma^2)}^{1/2}} e^{-\frac{r^2}{4\sigma^2}} e^{i(k_x x+k_y y)},\;\;\;l=1,2 
\end{equation} 
where $r^2=x^2+y^2$, $A_1=1/A$, $A_2=e^{i\phi}/A$ with $A = \sqrt
{A_1^2+A_2^2}$.  The rectangular box potential of height $V_0$ and
width $D$ is defined as follows:
\begin{equation}
  V(x)  = \left\{ \begin{array}{ll}
      V_0,  & \mbox{if  $0<x<D$}, \\
      0,       & \mbox{elsewhere}.
\end{array}
\right.
\end{equation} 

Given the linearity of the Dirac equation and the fact that
wavepackets are constituted by a Gaussian superposition of plane
waves, it is natural to express the transmission coefficient of a
Gaussian wavepacket of size $\sigma$ through the following
convolution:
\begin{equation}
\label{eq:TGAUSS0}
T_{\sigma}(k_x, k_y) = \int_{S_f} G \left( \frac{\vec{k}-\vec{k}'}{\sigma_k}
\right) T(k_x', k_y') dk_x' dk_y'
\end{equation}
where $S_f = \pi k_F^2$, with $k_F^2 = k_x^2+k_y^2$, denotes the Fermi
area, and $G$ a Gaussian kernel of width $\sigma_k = 1/\sigma$ in
mometum space. The function $T(k_x,k_y)$ is the transmission
coefficient of a plane wave with vector $\vec{k} \equiv (k_x, k_y)$,
which according to Ref. ~\cite{NATURE}, can be calculated as $T =
1-|r|^2$ with
\begin{equation}
\label{eq:TAN}
r=\frac{  2ie^{i\phi} (ss')^{-1} \sin(q_xD) (\sin\phi-ss' \sin\theta) }
{[e^{-iq_xD} \cos(\phi+\theta)+e^{iq_xD} \cos(\phi-\theta)]-2i
  \sin(q_xD)} ,
\end{equation}
being $\phi$ the incidence angle, $q_x^2 = (E-V_0)^2/\hbar^2 v_F^2$,
$\theta=tan^{-1}(k_y/q_x)$ the refraction angle, $s=\text{sign}(E)$,
$s'=\text{sign}(E-V_0)$, and $E$ the Fermi energy.

Since the transmission coefficient for a plane wave only depends on
the wave number $k_y$, and due to the fact that the $x$ component of
the wave vector experiences a perfect transmission, as a first-order
{\it approximation}, we perform the convolution in just one dimension,
$k_y$, that is:
\begin{equation}
\label{eq:TGAUSS}
T_{\sigma}(k) = \int_{-k_F}^{k_F} G \left( \frac{k-k'}{\sigma_k}
\right) T(k') dk'
\end{equation}
where we have defined $k \equiv k_y$. By setting $k'=k+q$, and
expanding $T(k+q)$ around $q=0$ to second order, Eq. \eqref{eq:TGAUSS}
delivers
\begin{equation}
\label{eq:TGAUSS2}
T_{\sigma}(k) \sim T(k) + \frac{\sigma_k^2}{2} T''(k) + O(\sigma_k^2)
\end{equation}
where $T''$ is the second derivative of $T$ with respect to $k$.  The
above expression means that resonant peaks ($T'(k_r)=0$, $T''(k_r)<0$)
are smoothed out whenever the filter width $\sigma_k$, is sufficiently
high, or, more precisely, $\sigma_k^2 > \frac{|T''(k)|}{2 T(k)}$.
This smoothing is the effect of non-resonant wavenumbers.  Given that
$\sigma = 1/\sigma_k$, one could readily estimate the minimal width
$\sigma$ above which the secondary resonant peak would no longer be
seen by the Gaussian wavepacket.  However, the asymptotic expansion
given by Eq. \eqref{eq:TGAUSS2} fails to represent the actual
transmission coefficient of the Gaussian wavepacket near the secondary
resonant peak, the reason being that, around that peak, a second order
expansion is grossly inaccurate because $\sigma^{2} T^{''} \sim 1$ and
higher orders will be even less accurate. As a result, the convolution
integral, Eq. \eqref{eq:TGAUSS}, needs to be computed.

\subsection{Computing the convolution}

\begin{figure}
  \centering   
  \includegraphics[scale=0.4]{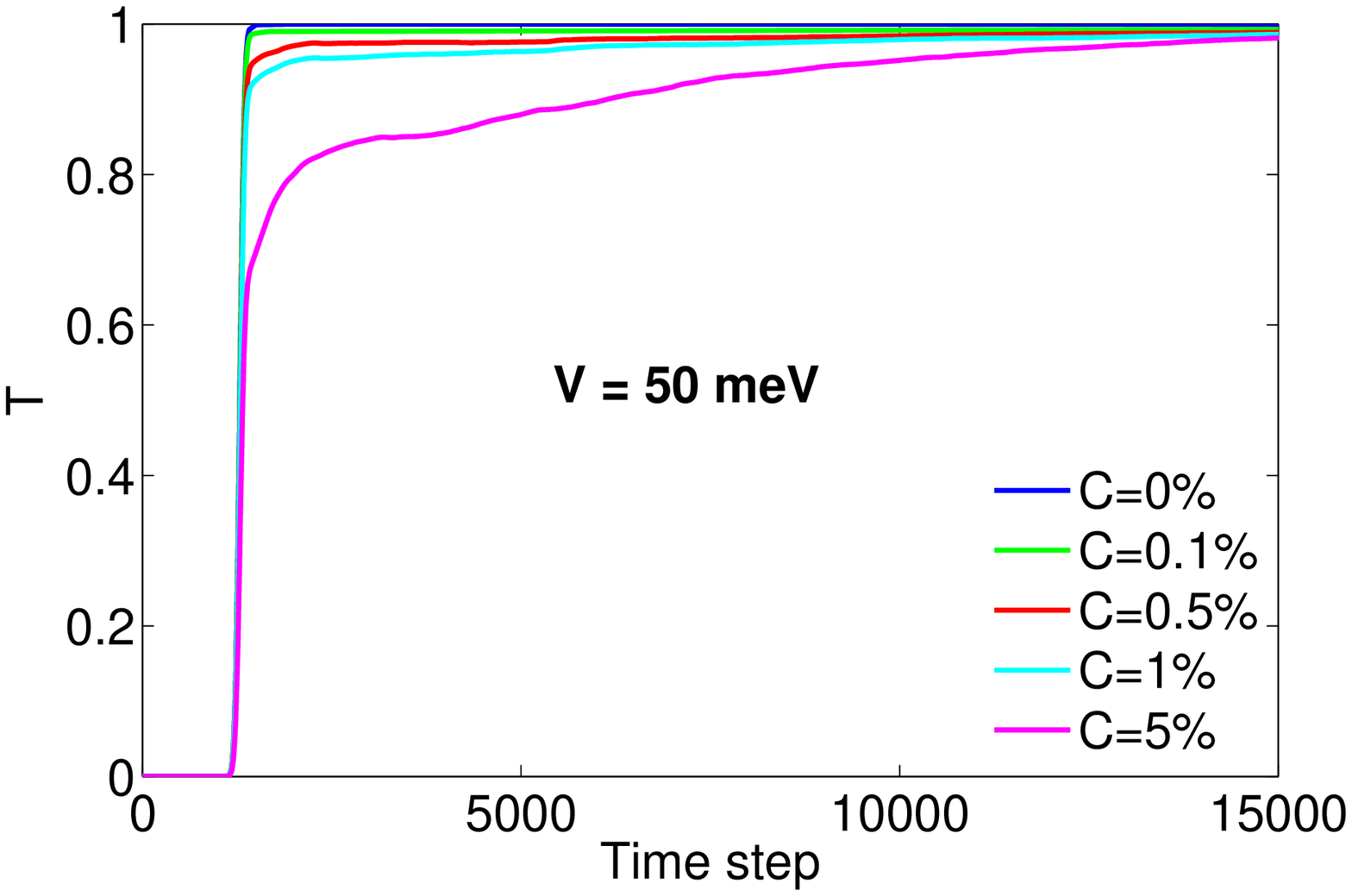}   
  \includegraphics[scale=0.4]{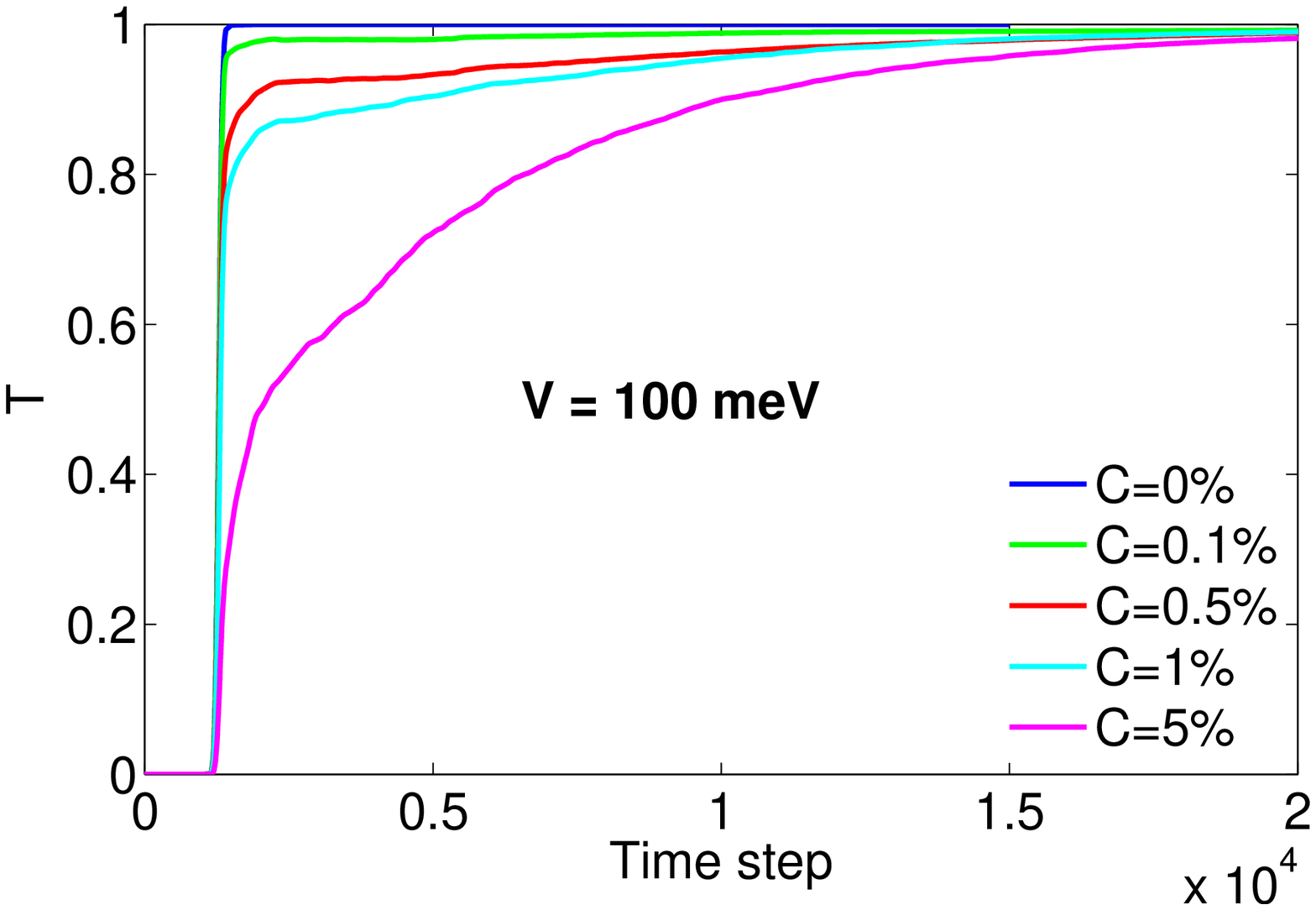}  \\   \vspace{0.2cm} 
 \includegraphics[scale=0.4]{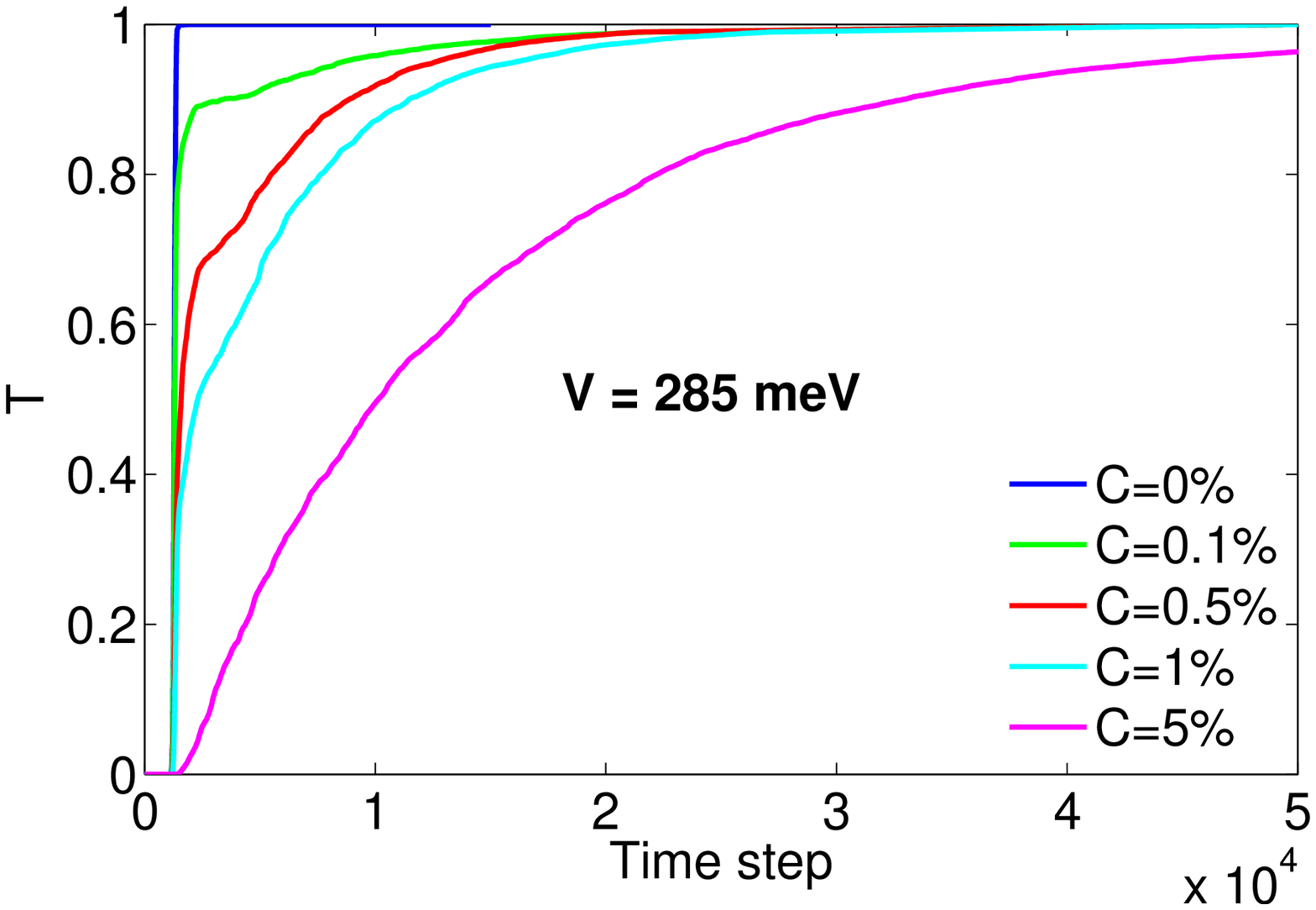}    
 \caption{Transmission coefficient as a function of time for the
   impurity potential set at $V = 50$, $100$, and $285$ meV while
   varying the impurity percentage $C$ ($C =$$0.1\%$, $0.5\%$, $1\%$
   and $5\%$) for $m=0$.}
  \label{fig:m0_fixed_v}
\end{figure}
To gain a quantitative sense of the dependence of the transmission
coefficient of the Gaussian wavepacket with the spatial spread
$\sigma$, we have numerically computed the convolution integral of
Eq. \eqref{eq:TGAUSS}, for the following values
$\sigma/D=0.15,0.31,0.46,0.92,1.85$, where $D=100$ nm is the width of
the potential barrier.  The parameters are the same as in Ref.
\cite{NATURE}, namely $E=0.08$ eV, $V_0=0.2$ eV and $D=100$ nm.  The
results are shown in Fig.\ref{fig:Fig2}.  From this figure, it is seen
that, for $\phi=0$, $T(k_r)=T(k_F cos(\phi_r))$ goes from $1$ to
$0.7348$, slightly over a $25$ percent reduction.  The same figure
also shows that around the secondary resonance (at $\phi =2 \pi/9$),
narrow wavepackets with $\sigma/D<0.46$ feature $T \sim 0.5$, with no
sign of the secondary resonant peak.  On the other hand, the secondary
peak is seen to re-emerge for $\sigma/D > 0.92$, i.e. when $\sigma$ is
of the order of $100$ nm, comparable with the barrier width.  With
$\sigma/D=1.85$, the secondary peak is recovered, but only to about
$80$ percent. Note that, for $\phi = \pi/2$, the transmission
coefficient is not zero, which is a consequence of the approximation
made to obtain Eq. \eqref{eq:TGAUSS} from Eq. \eqref{eq:TGAUSS0}.
However, as shown in Sec. \ref{numsimgauss}, the numerical simulation
of the transmission coefficient using QLB, shows generally a pretty
satisfactory agreement with the approximation Eq. \eqref{eq:TGAUSS}.

In order to use the plane-wave approximation, one needs to ensure that
the condition $\sigma>D$ is fulfilled, which sounds pretty
plausible. However, this condition is strongly dependent on the angle
of incidence.  In particular, it is far more stringent for oblique
than for head-on ($\phi=0$) incidence.  Indeed, for $\phi=0$,
$\sigma/D \sim 0.5$ yields a substantial $T=0.9$ for $\phi=0$, while
at $\phi=2 \pi/9$, we obtain a mere $T \sim 0.4$.  At $\sigma/D \sim 2
$, perfect transmission, $T=1$, is practically recovered at $\phi=0$,
while for $\phi=2 \pi/9$, $T \sim 0.8$, i.e. about $80\%$ percent of
full transmission.

We conclude that, for head-on incidence ($\phi \sim 0$), the
transmission coefficient of Gaussian packets is still similar to the
one of plane waves, as soon their extent becomes comparable to the
barrier width.  On the other hand, the secondary resonance, at oblique
incidence, is highly affected by the finite-size of the wavepacket,
and full recovery of perfect transmission seems to require wavepacket
extents significantly larger than the barrier width.

\subsection{Numerical simulations}
\label{numsimgauss}

The analytical expression of Eq. \eqref{eq:TGAUSS} has been compared
against direct numerical simulation of the Dirac equation, using the
quantum lattice Boltzmann (QLB) method.  In order to back-up the
previous findings, we have computed full numerical solutions of the
Dirac equation using a quantum lattice Boltzmann solver.  The
simulations are performed on $1024^2, 2048^2$ and $4096^2$ grids,
depending on the size of the Gaussian packet.  Lattice units are
chosen such that $\tilde D=D/\Delta x=52$, $dt=dx/v_F=1.92 \times
10^{-15}$ seconds and energy is normalized in units of $\hbar/dt$.
The physical parameters are taken from Ref.  \cite{NATURE}, that is
$E=0.080$ eV, $V_0=0.200$ eV and $D=100$ nm.  The following sequence
of wavepackets spreading, $\sigma=24,48,96$ has been simulated, with
$D=52$, all in lattice units.  The results of the QLB simulations
appear substantially in line with the prediction of the convolution
integral, i.e. they clearly show the disappearance of the secondary
peak for $\sigma/D<0.46$, and its progressive reappearance above this
threshold (see Fig.\ref{fig:Fig3}). Note that, different from the
solution of the convolution integral, Eq. \eqref{eq:TGAUSS}, the
transmission coefficient measured by the simulation is zero for
$\phi=\pi/2$, as should be expected, but the appearance of the second
resonant peak is still retained.

In Fig. \ref{fig:Fig5}, we show typical snapshots of the wavepackets
for the cases $\phi=0$, $2 \pi/9$, and $\pi/3$, for $\sigma/D=1.85$.
The snapshots clearly show that, in the case $\phi=0$, the wavepacket
crosses the barrier totally unperturbed, with literally no distortion
at any stage of the evolution.  In the case of oblique resonant
propagation, the packet still manages to cross the barrier to a large
extent, ($T=0.76$), with significant distortions in the intermediate
stages of the evolution, leaving $24$ percent of the packet
behind. Finally, in the case of oblique non-resonant propagation,
$\phi=\pi/3$, the packet is mostly bounced-back by the barrier, with a
transmission coefficient as low as $T=0.13$.

\section{Klein paradox in random media}

\begin{figure}
  \centering   
  \includegraphics[scale=0.4]{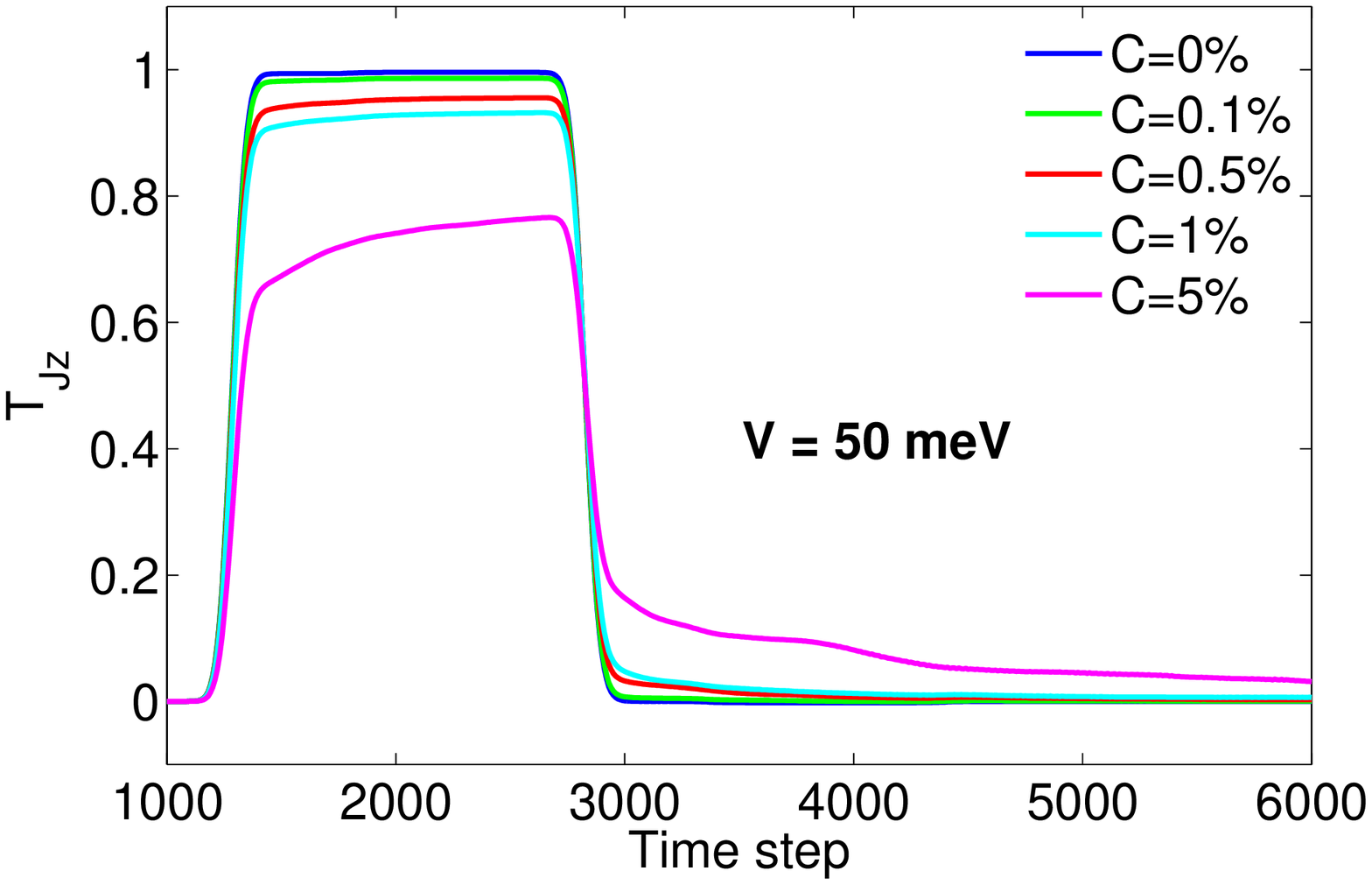}   
  \includegraphics[scale=0.4]{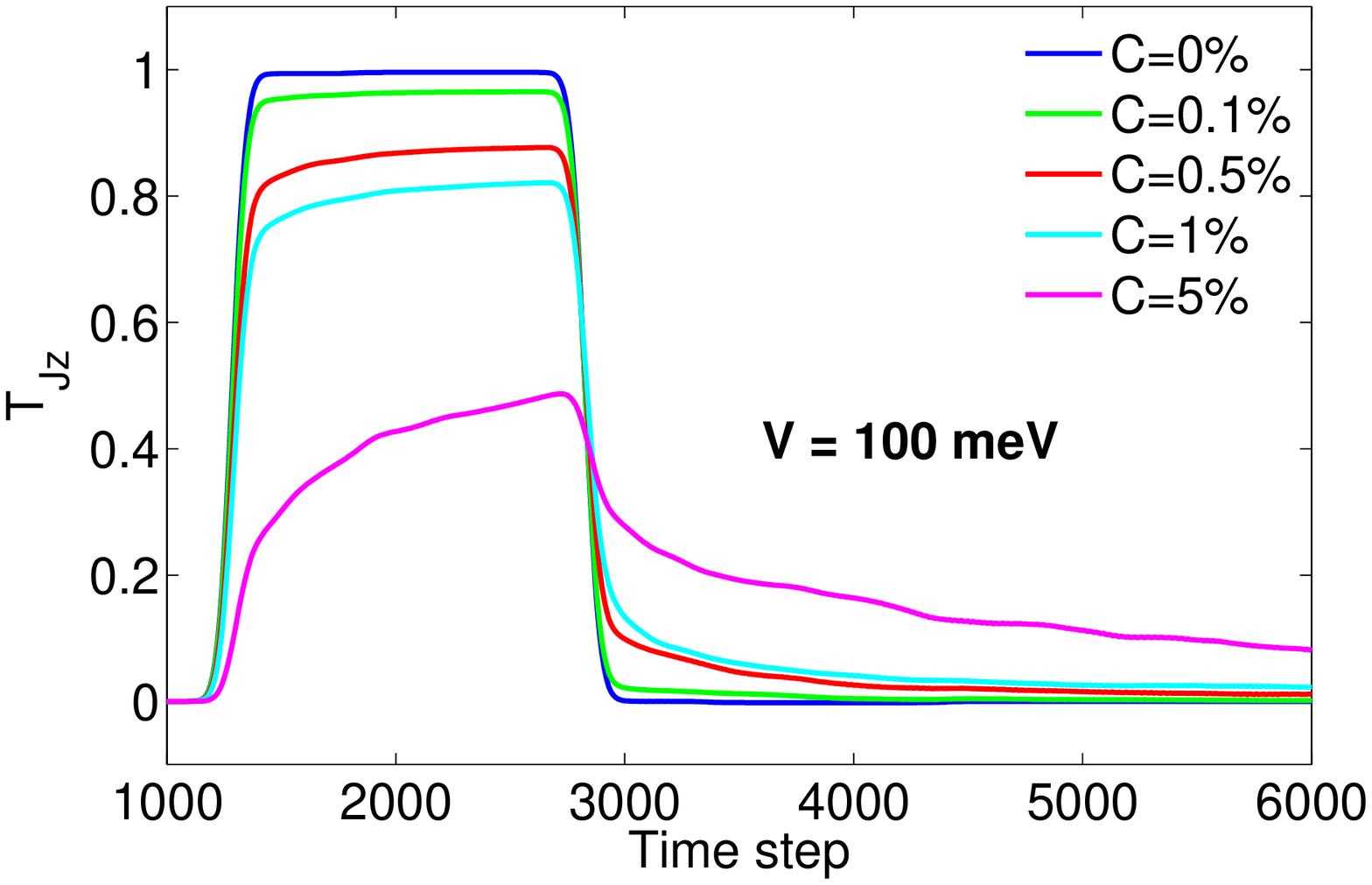}  \\   \vspace{0.2cm} 
  \includegraphics[scale=0.4]{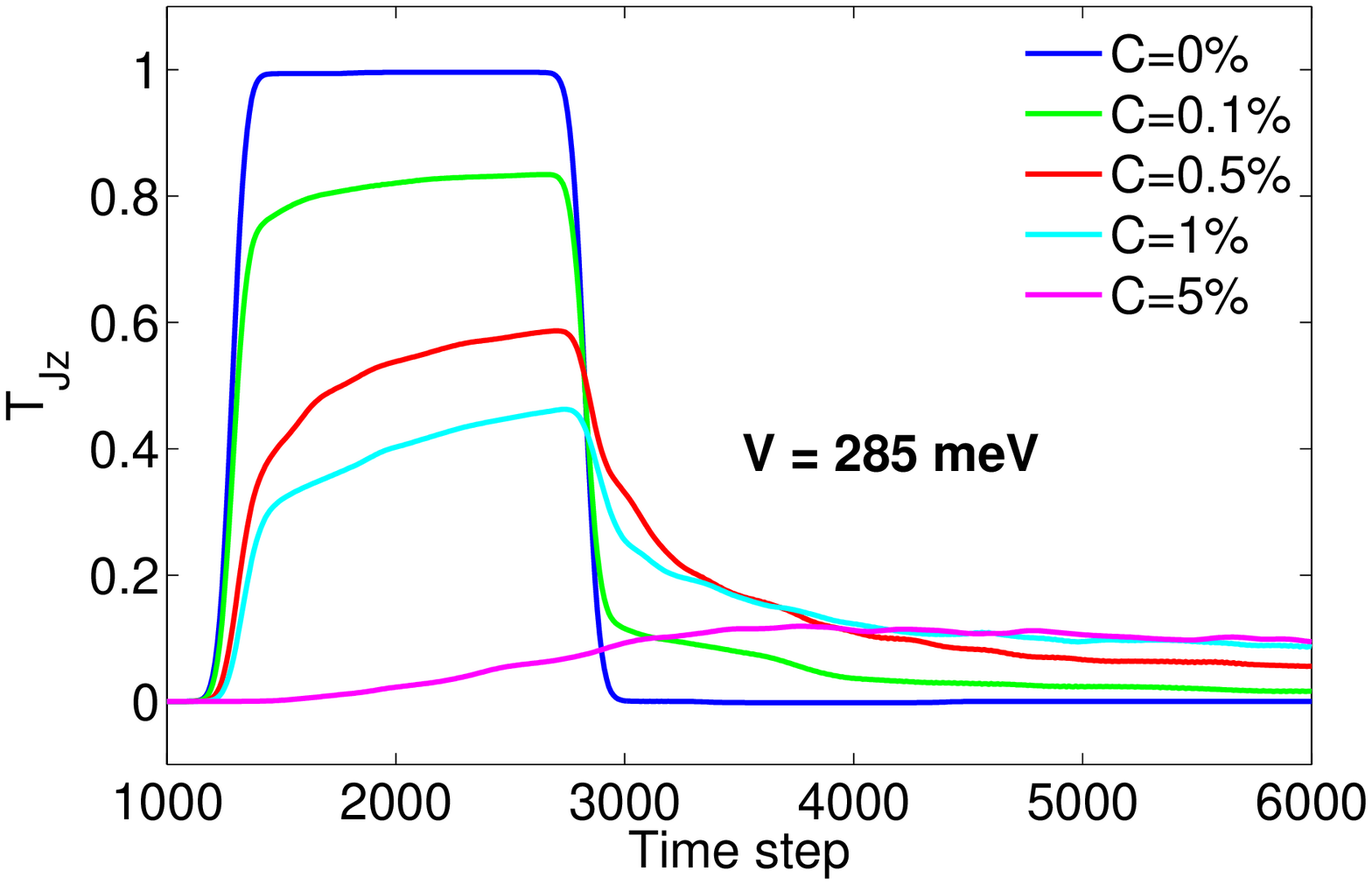}    
  \caption{Momentum transmission coefficient $T_{Jz}$ as a function of
    time for the impurity potential set at $V = 50$, $100$, and $285$
    meV while varying the impurity percentage $C$ ($C=$$0.1\%$,
    $0.5\%$, $1\%$ and $5\%$) for $m=0$.}
  \label{fig:m0_jz_v}
\end{figure}

\begin{figure}
   \centering   
   \includegraphics[scale=0.4]{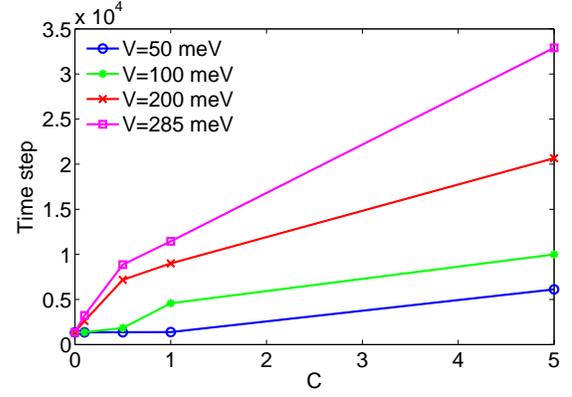}  \\   \vspace{0.2cm} 
   \caption{Time at which $90\%$ of the wave packet has been
     transmitted, $t_{0.9}$, as a function of the impurity percentage
     for fixed values of $V$ and $m=0$.  The potential barriers are as
     follows: $V=50$, $100$, $200$ and $285$ meV. The impurity
     percentage values are $C=$$0.1\%$, $0.5\%$, $1\%$ and $5\%$.  }
   \label{fig:timeStepT09}
\end{figure}

\begin{figure}
  \centering   
  \includegraphics[scale=0.15]{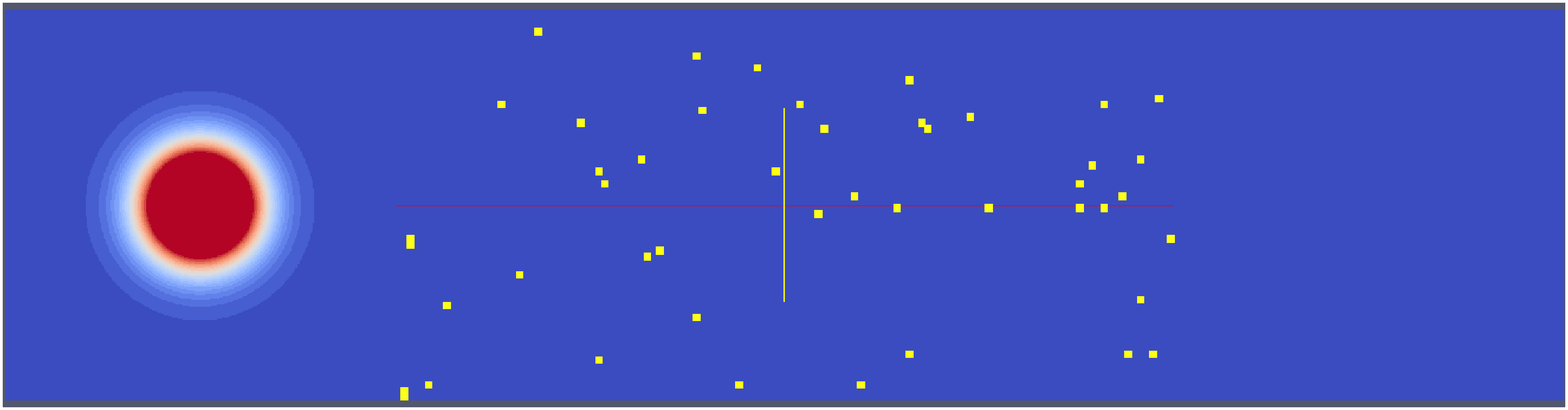}  
  \includegraphics[scale=0.15]{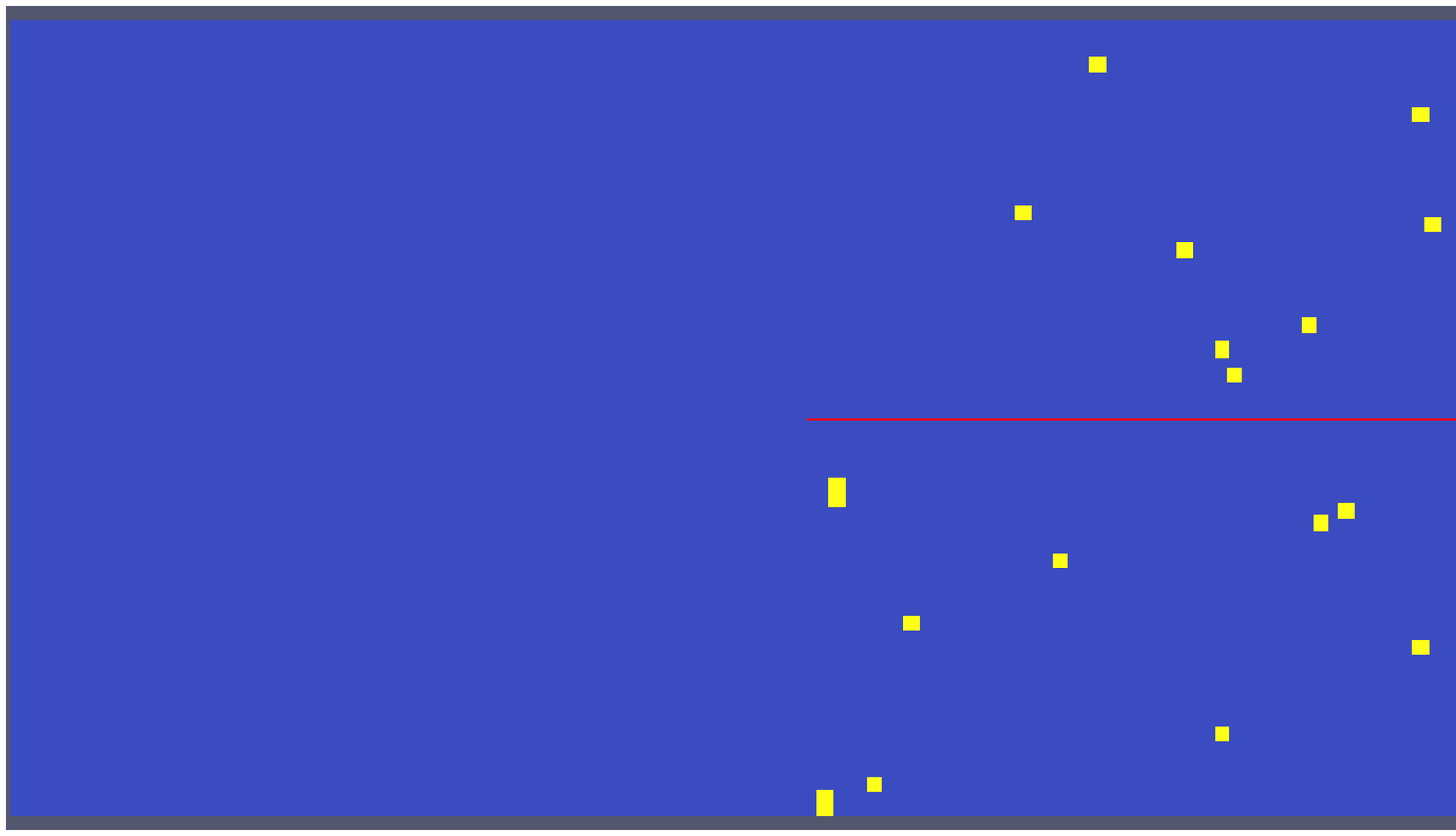}   
  \includegraphics[scale=0.15]{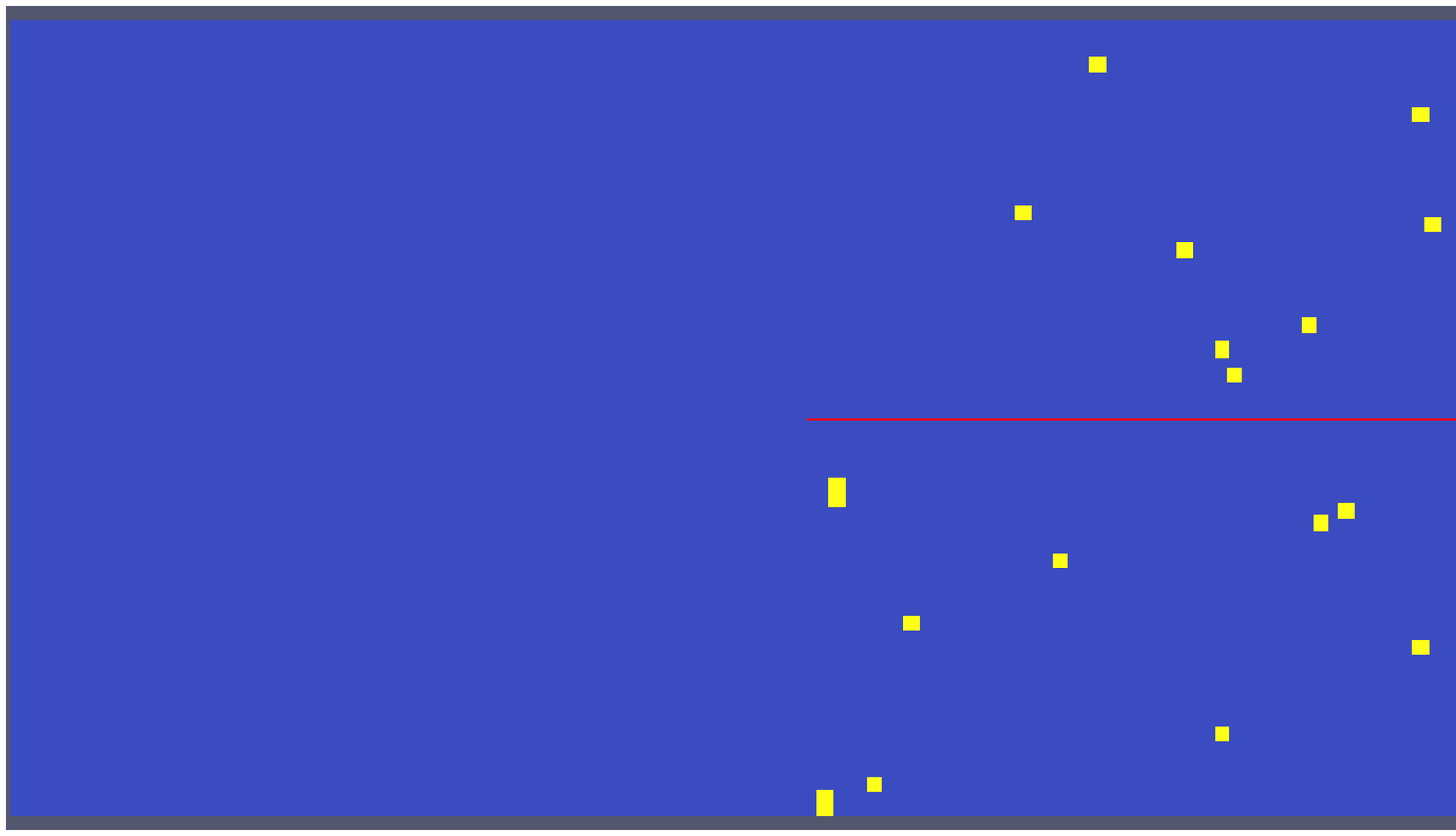} 
  \includegraphics[scale=0.15]{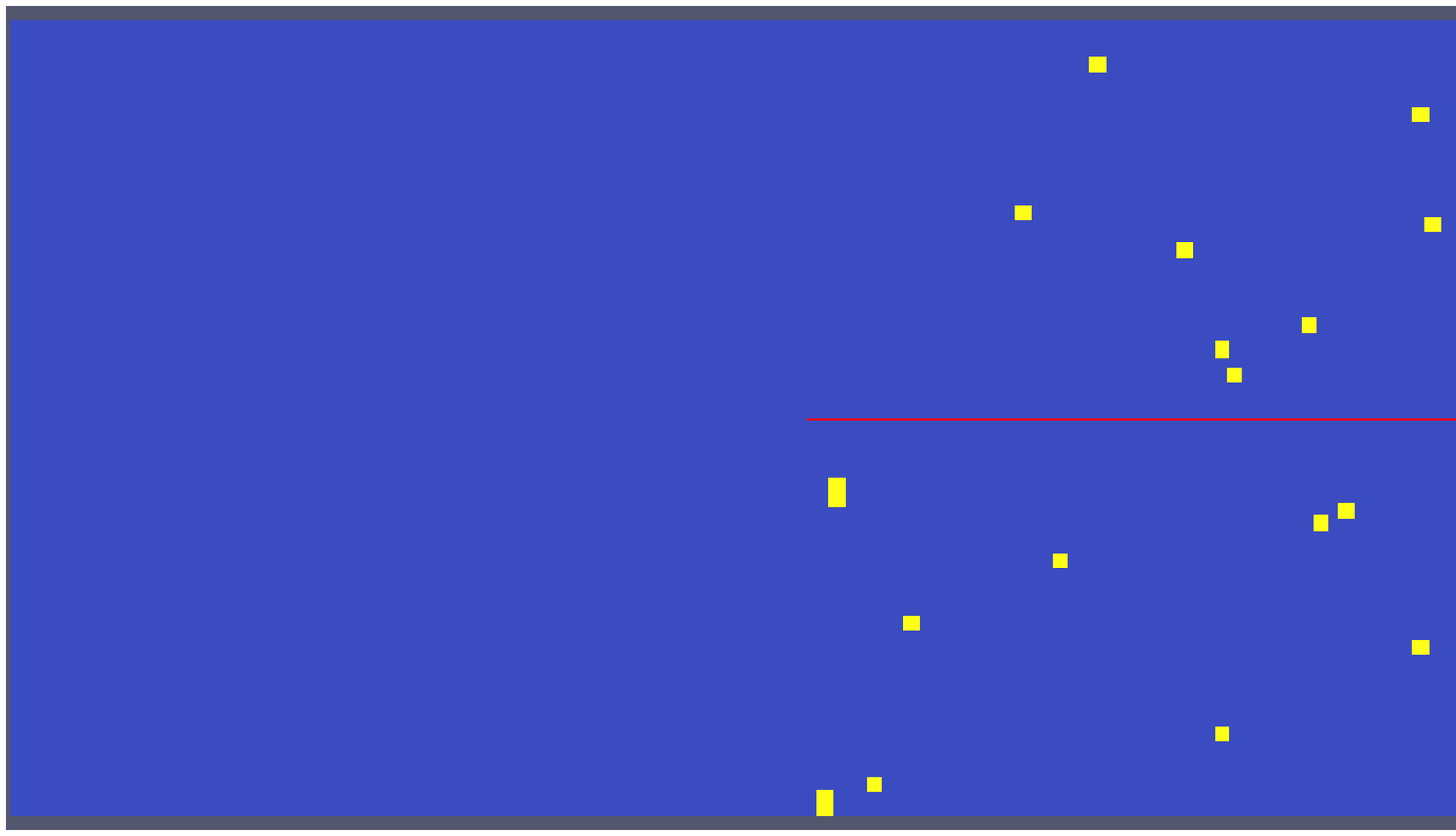} 
  \caption{Wave packet density $\rho$ at times = $0$, $900$, $1500$,
    and $1800$ (lattice units) for the simulation performed with
    impurity percentage $C=0.5\%$ and $V = 50$ meV.}
  \label{fig:imp05_v50}
\end{figure} 

One of the major technological challenges in current graphene research
is to manufacture larger samples, above $10$ microns, for practical
use in engineering devices \cite{ENG}. As the sample size is
increased, however, it becomes more and more difficult to secure the
purity of the sample, i.e. avoid crystalline inclusions (impurities)
which alter the local structure of the graphene honeycomb lattice.
Such impurities are indeed known to significantly affect the
macroscopic properties of the sample, primarily its electrical
conductivity. To gain insight into this problem, it is therefore of
interest to investigate the propagation of relativistic wavepackets
within a disordered sample.  

The conductivity of two-dimensional massless fermions in disordered
media has made the object of intense studies in the literature,
\cite{MFD}.  The contribution of the present work to this subject
relates to the following three directions, i) Investigate the
Klein-Paradox for the case of Gaussian wave-packets rather than plane
waves, both for single barriers and disordered samples, ii) Discuss
the viability of semi-classical descriptions of electrons excitations
in disordered media, based on quantitative analogies with flows in
porous media, iii) Expose the quantum lattice Boltzmann method as a
new computational tool for electron transport in graphene, which might
bear a special interest for prospective implementations on parallel
computers.  Notwithstanding points i-iii) above, we wish to point out
that, being our solution based on the single-particle Dirac equation
(no many-body effects), any conclusion on transport phenomena in
actual graphene samples must be taken with great caution.
We also wish to remark that the Klein tunneling is expected to be
relatively mild in the present set up, for two reasons.  
First, because the Gaussian wavepacket always includes non-resonant
frequencies suffering partial reflection; second, because, being the
wavepacket wider than the obstacle size (see below), it can split and
turn around the obstacle like a classical fluid, hence be partially
transmitted, without any quantum tunneling through the barrier.
 
To analyze these transport phenomena, we simulate the propagation of a
relativistic Gaussian wavepacket through a two-dimensional domain
composed of three regions: an inlet region, where the wave packet is
positioned at the initial time $t=0$; the impurity region, i.e. the
central part of the domain where randomly distributed barriers
(impurities) are located; and the outlet region, which is the final
region, where measurements of the transmitted wave packet are
taken. Due to the large effective fine structure constant in graphene,
we will neglect in our study the Coulomb interaction between
carriers. The impurity concentration is given by $C=Nd^2/A$, where $N$
is the number of square obstacles of cross section $d^2$, distributed
over an area $A=L_y \times L_z$.  For the present simulations $d=8$
(larger than the typical lattice distante of graphene) and $C$ is
varied in the range $0.001 \div 0.05$.  In Fig. \ref{fig:domain}, the
computational domain is sketched, periodic boundary conditions are
imposed at top and bottom boundaries, while a bounce-back condition is
enforced at the inlet , and an open boundary condition is imposed at
the outlet (so that the transmitted wave packet is not reflected
back). We use a square lattice of size $2048\times 512$ cells, such
that the regions $[0, 512)\times 512$, $[512, 1536) \times 512$, and
$[1536, 2048] \times 512$ correspond to the inlet, impurity, and
outlet regions, respectively. The cell size is chosen to be $\Delta x
= 0.96$nm, and the spreading of the initial Gaussian wave packet
$\sigma = 48$ (in lattice units), leading to a Fermi energy $E_F =
0.117$ ($80$meV in physical units). In our study, we use two values
for the mass of the particles, $m=0$ (ungaped graphene) and $m=0.1$
(gaped graphene), and vary the impurity potential and the
concentration.  Five barrier heights are considered, namely
$V=25,50,100,200,285$ meV.  Note that, while the first two lie below
$E_F$, hence can be overcome classically, the others can only be
traversed head-on via quantum tunnelling.  It should be further
observed, though, that since the wavepacket is wider than the single
impurity, i.e. $\sigma>d$, even in the case $E_F<V$, the wavepacket
can split and turn around the obstacle like a "classical" fluid.  Our
results can be classified according to the energy of the particles,
the potential of the barrier, and their mass as follows: weak
potentials, $V < E_F - mv_F^2$; intermediate potentials, $E_F - mv_F^2
< V < E_F + mv_F^2$; and strong potentials, $V > E + mv_F^2$.  The
transmission coefficient $T(t)$ is obtained by computing $T(t) =
\int_{z>z_{outlet}} \rho(z,y,t) dz dy$, where $\rho$ is the wave
packet density defined as $\rho = |u_1|^2 + |u_2|^2 + |d_1|^2 +
|d_2|^2$, with $\psi=(u_1,u_2,d_1,d_2)^T$ being the Dirac
quadrispinor.

\subsection{Wave packet mass $m=0$}

\begin{figure}
  \centering   
  \includegraphics[scale=0.4]{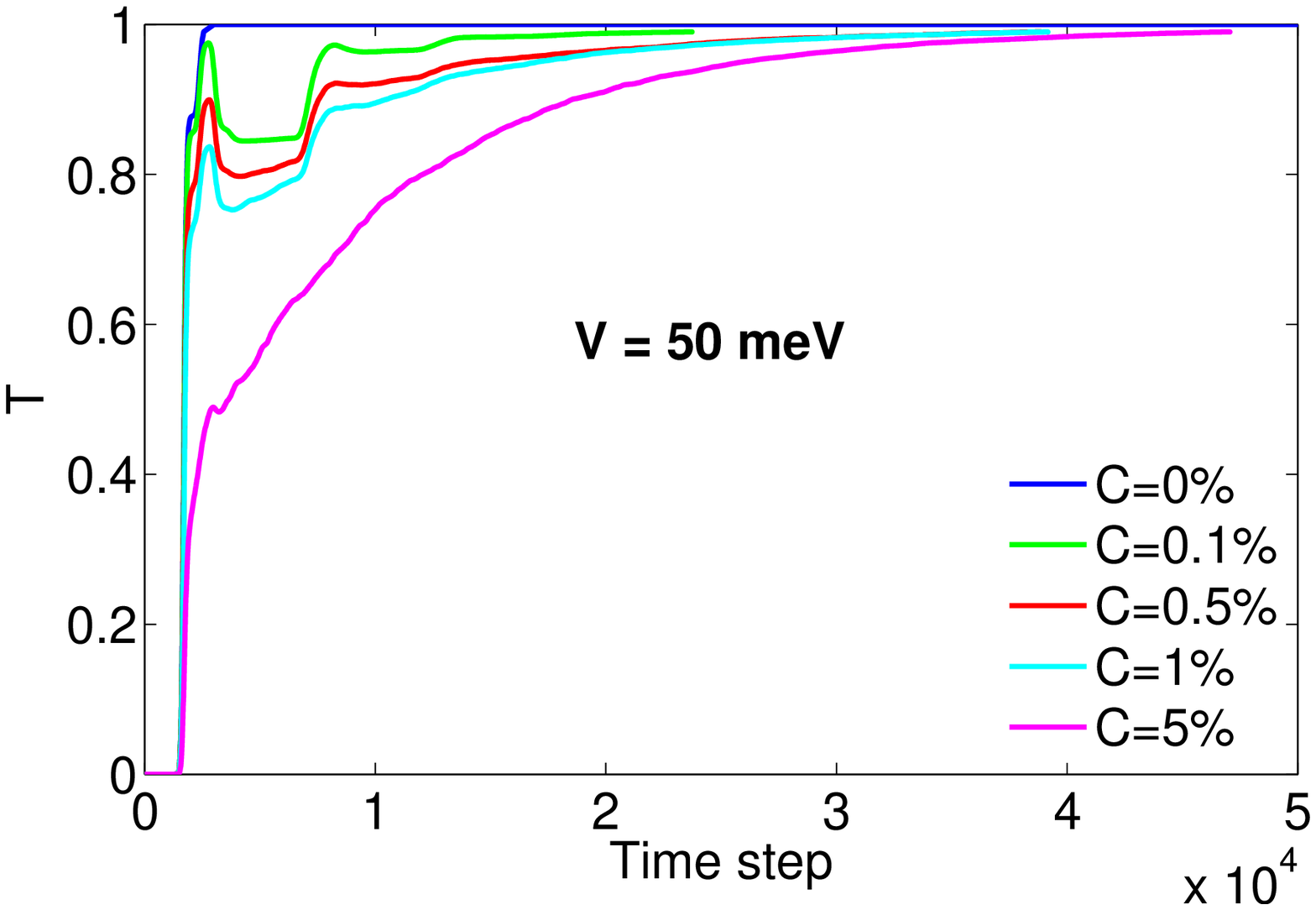}   
  \includegraphics[scale=0.4]{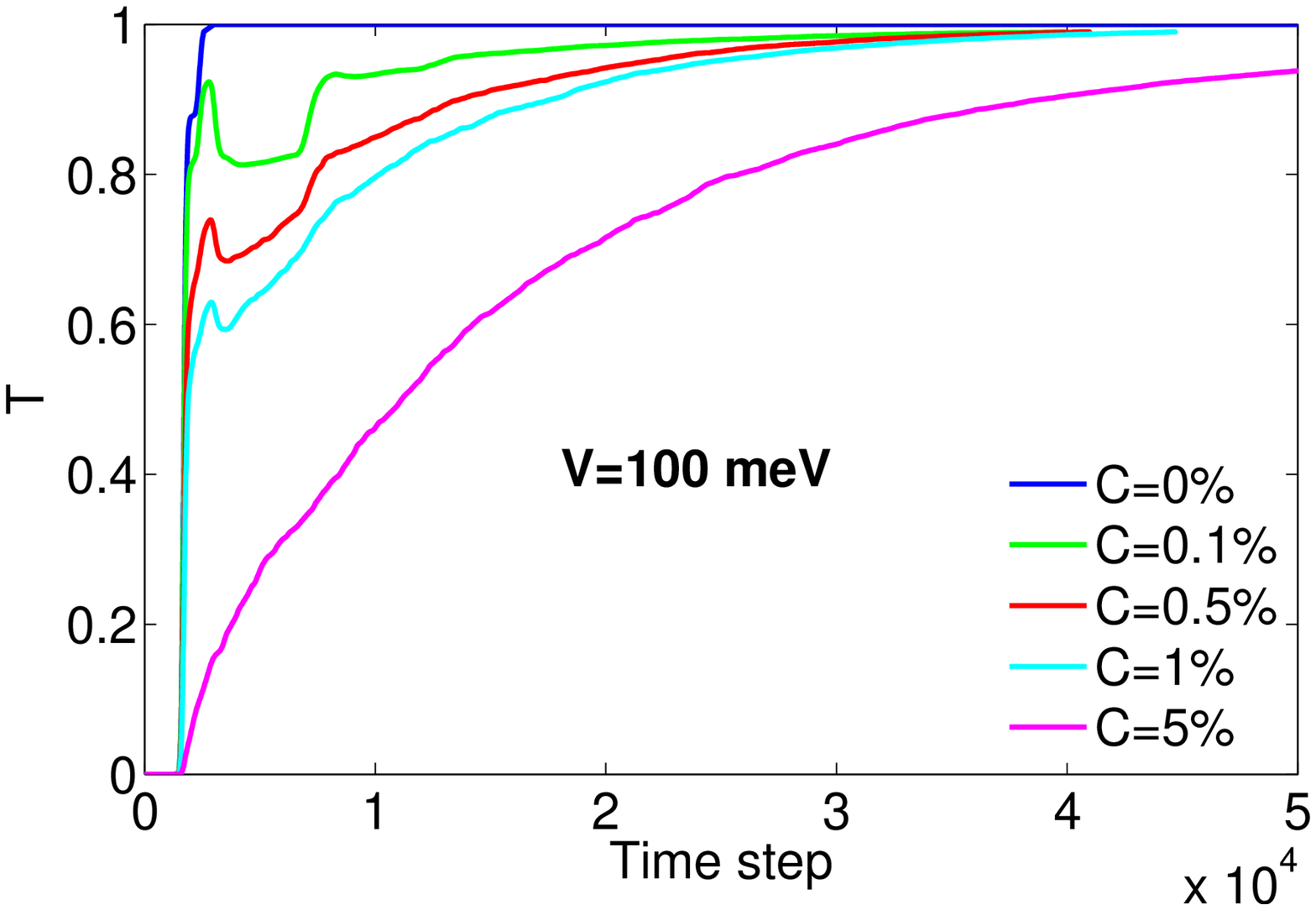} \\ \vspace{0.2cm}
  \includegraphics[scale=0.4]{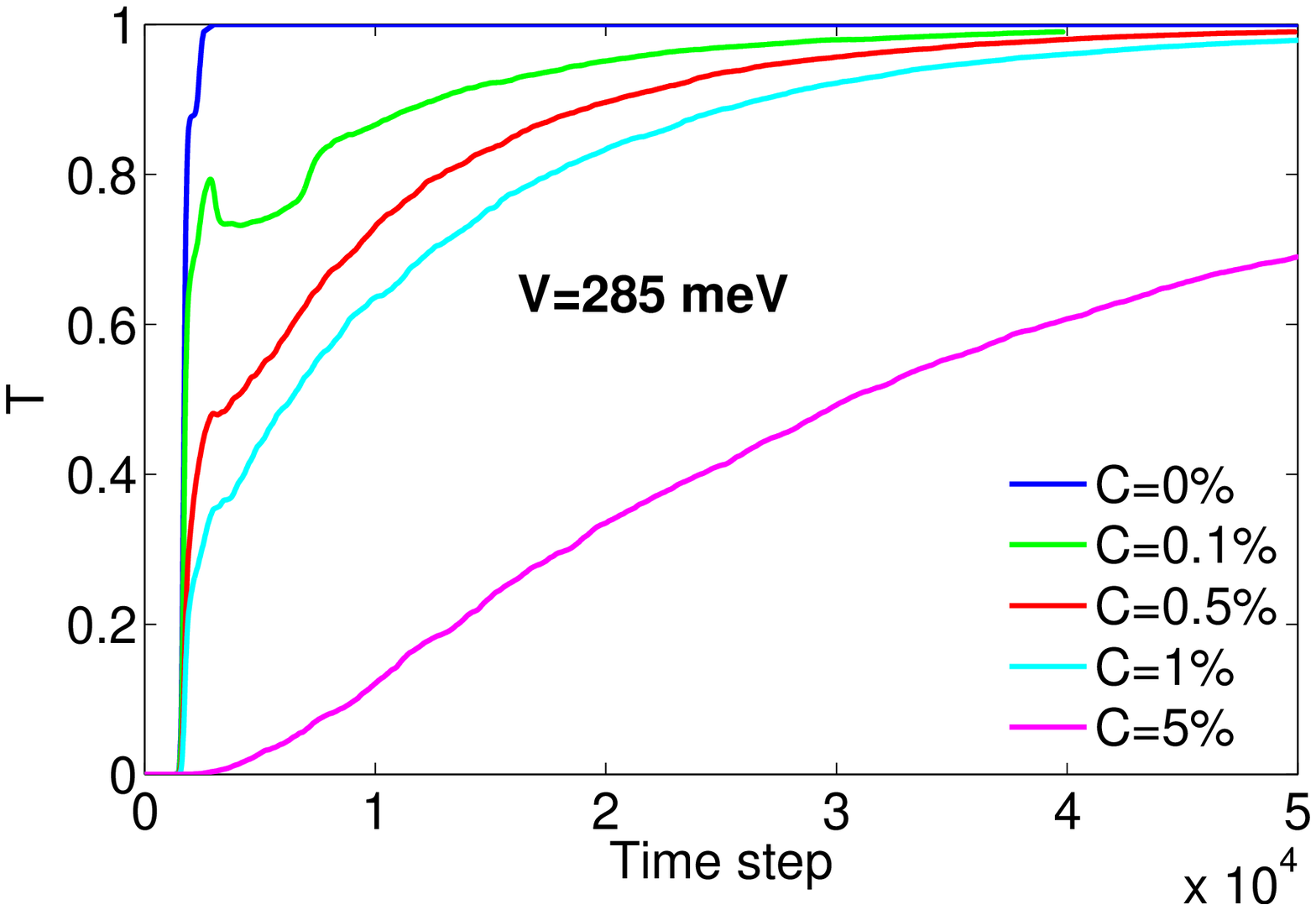}    
  \caption{Transmission coefficient as a function of time for the
    impurity potential set at $V = 50$, $100$, and $285$ meV while
    varying the impurity percentage ($C=0.1\%$, $0.5\%$, $1\%$ and
    $5\%$) for $m=0.1$.}
  \label{fig:m1_fixed_v}
\end{figure}

\begin{figure}
  \centering   
  \includegraphics[scale=0.4]{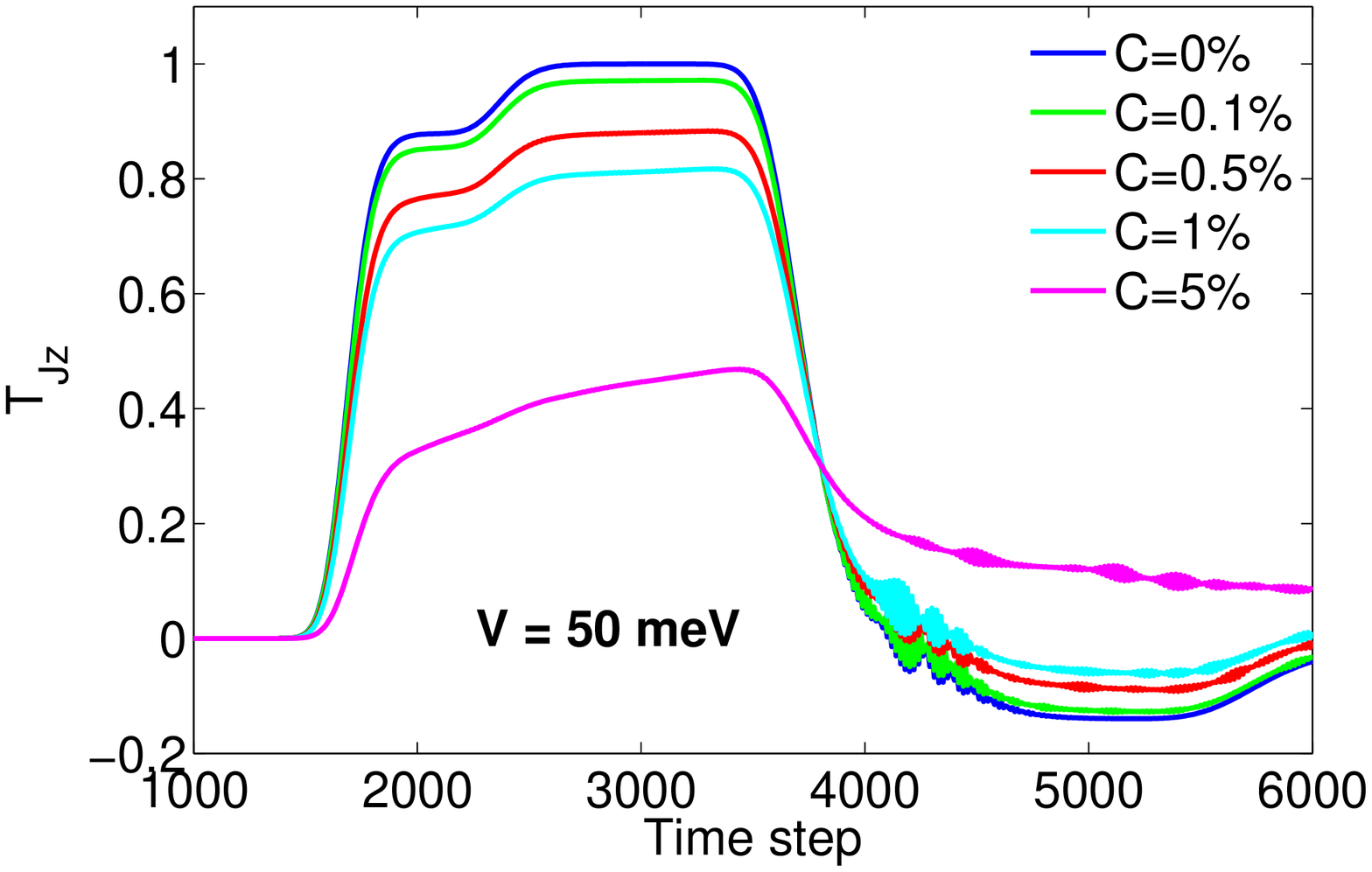}   
  \includegraphics[scale=0.4]{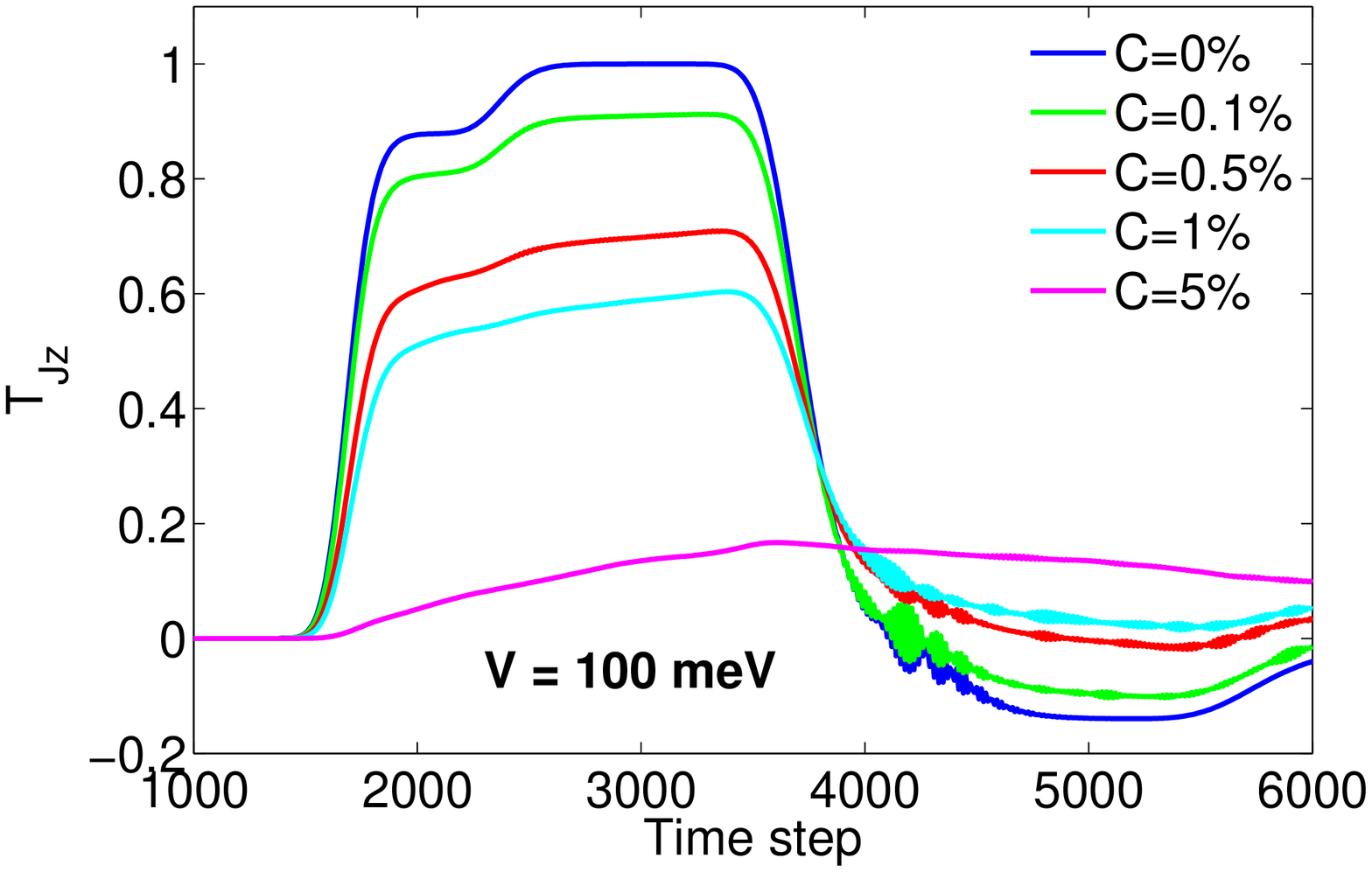}  \\   \vspace{0.2cm} 
  \includegraphics[scale=0.4]{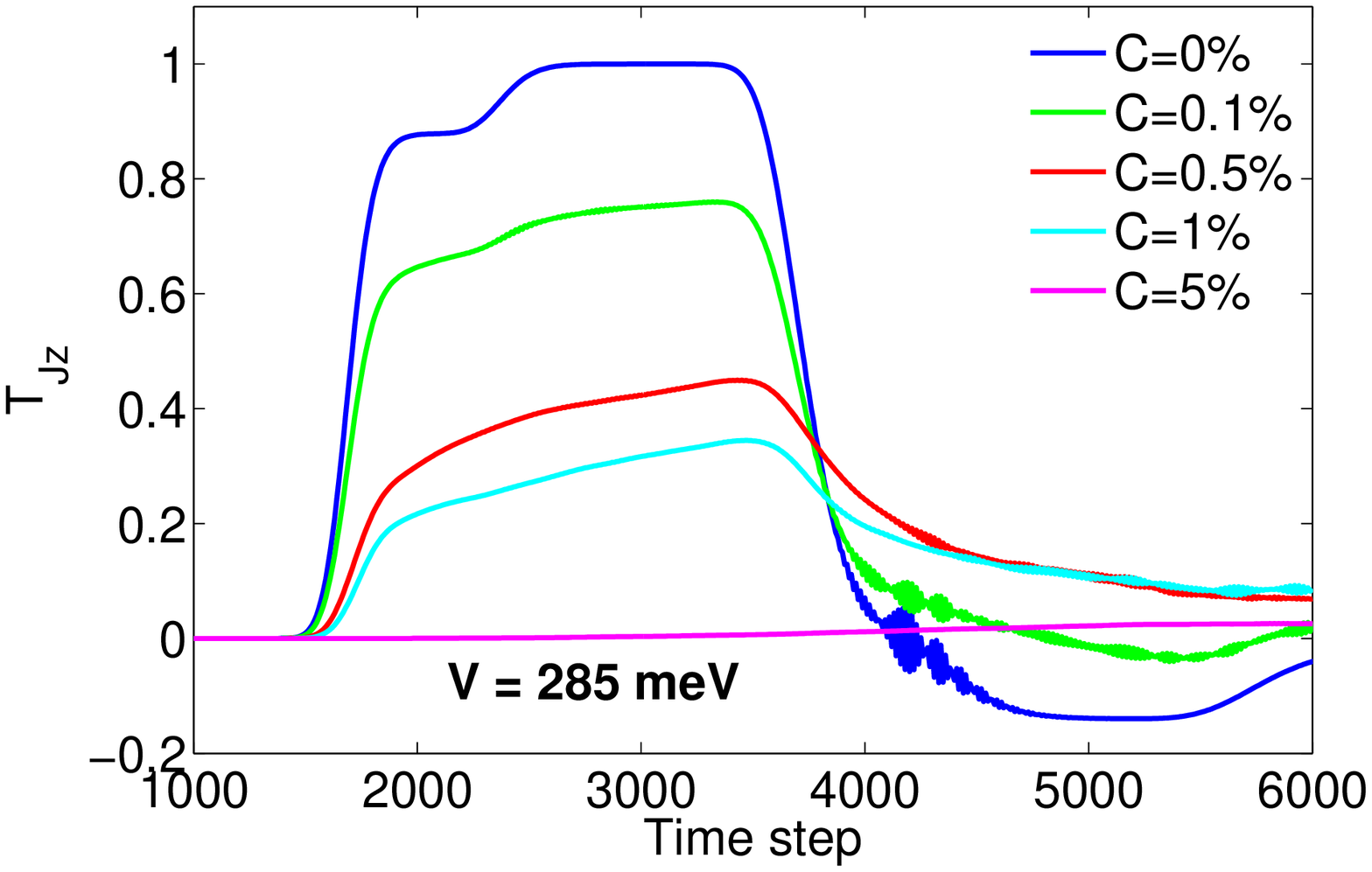}    
  \caption{Momentum transmission coefficient $T_{Jz}$ as a function of
    time for the impurity potential set at $V = 50$, $100$, and $285$
    meV while varying the impurity percentage ($C$ = $0.1\%$, $0.5\%$,
    $1\%$ and $5\%$) for $m=0.1$.}
  \label{fig:m1_jz_v}
\end{figure}

\begin{figure}
  \centering   
  \includegraphics[scale=0.15]{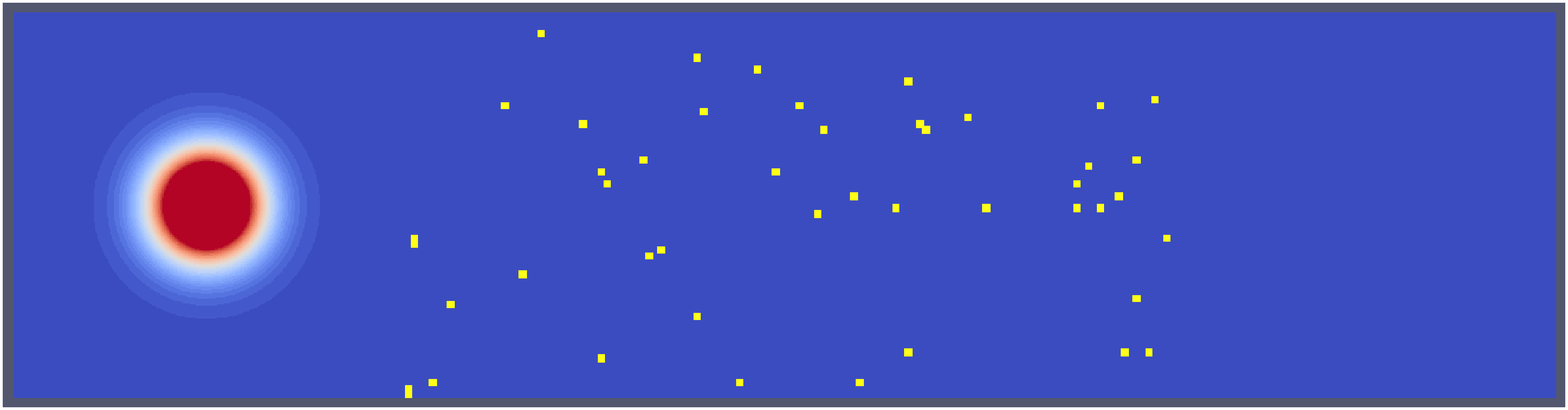}   
  \includegraphics[scale=0.15]{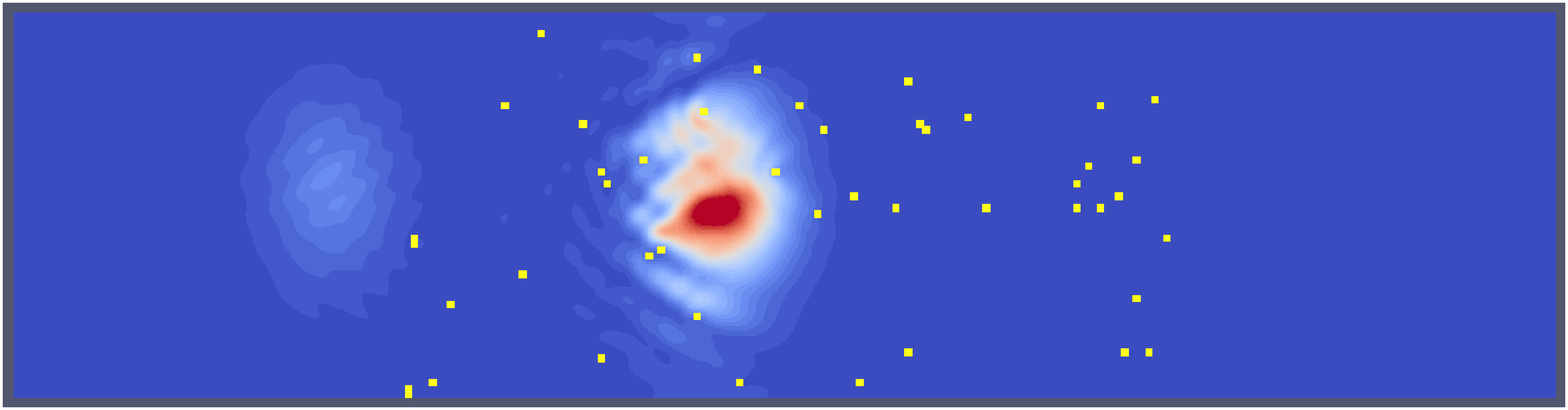}  
  \includegraphics[scale=0.15]{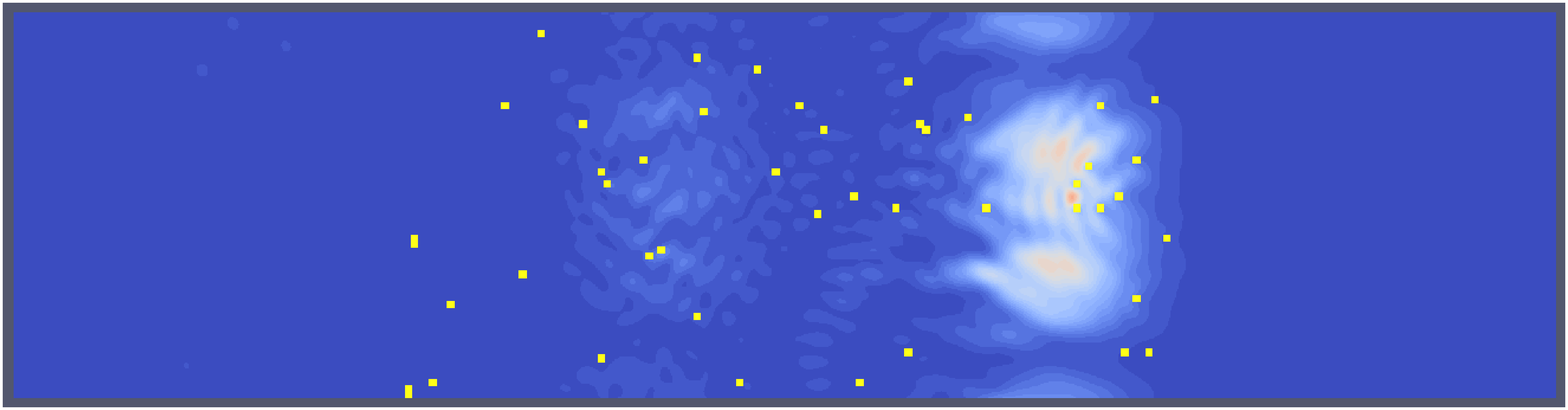}   
  \includegraphics[scale=0.15]{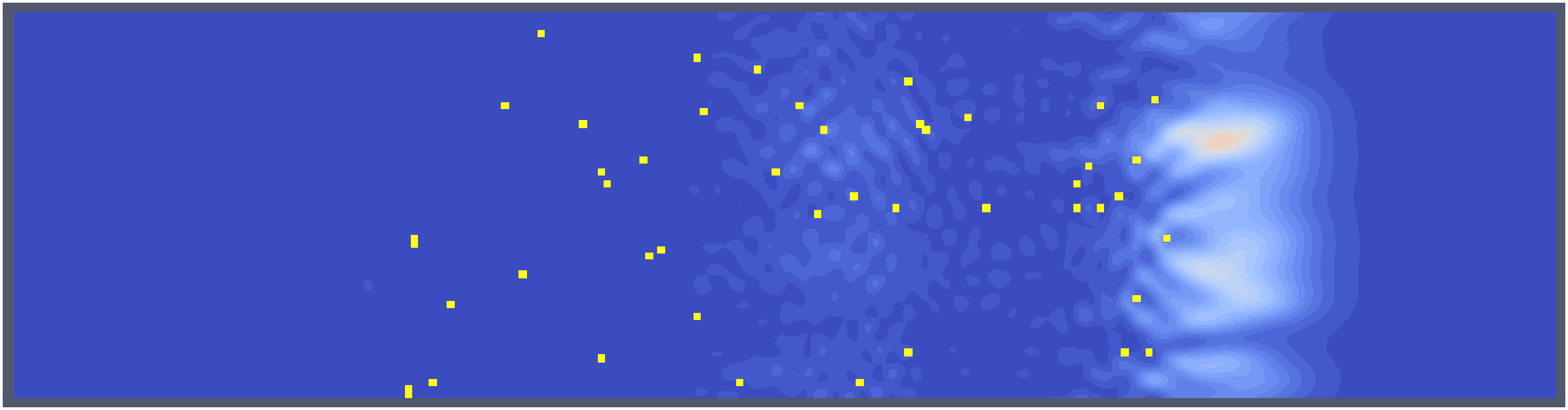}   
  \caption{Wave packet density $\rho$ at times $0$, $900$, $1500$, and
    $1800$ (lattice units) for the simulation performed with impurity
    percentage $=0.5\%$ and $V = 50$ meV and with $m=0.1$.}
  \label{fig:imp05_v50_m1}
\end{figure}

In this first set of simulations, we fix $m=0$, and vary the impurity
concentration, $C$, and the strength of the impurity potential, $V$.
In Fig.~\ref{fig:m0_fixed_v}, we fix the value of $V$ and we compare
$T$ while varying the impurity percentage, including the reference
value for the pure sample $C=0$. From this figure, we observe that the
wave packet takes longer to regroup for high impurity concentration
and high impurity potential. This is a natural consequence of the
randomness induced in the wavefunction by the disordered
media. However, in all cases, the complete wave packet is
reconstructed after some time, with no stagnant regions left behind.
This can be related to the momentum loss due to the presence of the
impurities, and therefore the motion of the wave packet experiences a
corresponding slow-down. Note that, in order to recover the complete
wave function, the simulations have been performed in a longer domain.
Otherwise the right-moving wave-packet would leave the outlet
region too early while the left-mover is still in the domain.  In order to
provide a measurement of momentum dissipation, i.e. the loss of
conductivity due to impurities, we compute the momentum transmission
coefficient as follows:
\begin{equation} \label{eq:TJz} 
T_{Jz}(t) = \int_{z>z_{out}} J_z(z,y,t) \quad dz dy,
\end{equation}
where
\begin{equation} \label{eq:Jz}
  J_z = \vec{\psi}^{\dagger} A_z \vec{\psi} + \vec{\psi}^{\dagger} A_z^{\dagger} \vec{\psi},
\end{equation}
is the $z$-component of the current density with $A_z$ the streaming
matrix along $z$ and $\vec{\psi}=(u_1,u_2,d_1,d_2)^T$ the Dirac
quadrispinor.

In Fig.~\ref{fig:m0_jz_v}, we fix the value of $V$ and compare
$T_{Jz}$, while varying the impurity percentage. The subscript $_{Jz}$
denotes the transmission coefficient due to the $z$-component of the
current density, $J_z$. As a reference, we also plot $T_{Jz}(t)$ when
the impurity percentage is set to $C=0$.  From Fig.~\ref{fig:m0_jz_v}
we can observe that, unlike the density, the momentum transmission
coefficient does not saturate at unity (its value in the inlet region
at the beginning of the simulation), because momentum is irreversibly
lost in the impurity region.  Furthermore, as expected, the momentum
loss increases with increasing impurity potential and concentration.

As a characteristic quantity associated with the dynamics of the
transmission coefficient $T$, in Fig.~\ref{fig:timeStepT09}, we report
the escape time, $t_{0.9}$, i.e. the time at which the transmission
coefficient reaches $90\%$, (i.e. at $90\%$ of the wave packet is
transmitted through the obstacle region). As above, we plot $t_{0.9}$
as a function of the impurity percentage for two values of $V$.  We
notice that for high impurity concentration the Gaussian wave packet
takes longer to cross the impurity barrier.  The same effect occurs
when the impurity potential is increased.  At low impurity
concentration, $C=0.001$, the effect of the potential barrier is
relatively minor, but, as the concentration is increased, the escape
time grows approximately linearly with the barrier voltage.

In Fig.~\ref{fig:imp05_v50}, we show some representative snapshots of
the first $1800$ time steps of the simulation, for impurity percentage
$C=0.5\%$ and $V=50$ meV. Here, we can see the way how the wave packet
is scattered by the impurities, generating a plane front, as a result
of the fragmentation of the wavefunction due to the random obstacles.

\subsection{Wave packet mass $m=0.1$}

Next, we repeat the same simulations for the case of massive
particles, with $m=0.1$.  Note that, since $mv_F^2/E_F=0.83$, the rest
energy is a significant fraction of the kinetic energy, and therefore
the wavefunction comes in the form of a superposition of two
wavepackets, both moving at the Fermi speed, along opposite
directions, and mixing through the non-zero mass term.

In Fig.~\ref{fig:m1_fixed_v}, we fix the value of $V$ and compare $T$,
while varying the impurity concentration $C$.  As a reference, we also
plot $T$ with $C=0$.  From the results, we observe that the wave
packet takes longer to cross the impurity region than for the case of
$m=0$ (the time it takes to reach a unit value of the transmission
coefficient is longer).  This is due the slow-down of the wavefunction
as compared to the Fermi speed, because of the non-zero particle mass.
Note the peak in the transmission coefficient, once the wave packet
exits from the impurity region. This is due to the fact that $T_{J_z}$
takes negative values in the late stage of the evolution, indicating
the prevalence of the left-moving component of the wavepacket once the
right-moving one has left the domain.

We compute the momentum transmission coefficient using equations
\eqref{eq:TJz} and \eqref{eq:Jz}.  In Fig.~\ref{fig:m1_jz_v}, we fix
the value of $V$ and compare $T_{Jz}$ while varying the impurity
percentage. As a reference, we also plot $T_{Jz}(t)$ when the impurity
percentage is set to zero.  Note that, as expected, due to the inertia
when the mass is increased, the curve of the momentum transmission
becomes wider than for the case of zero mass, reflecting the fact that
the wave packet takes longer to move across the impurity region. In
addition, the maximum momentum is smaller than for the case of zero
mass, which indicates higher momentum losses. Thus, a non-zero mass of
the (quasi)-particles, results in higher momentum losses.  Also to be
noted, are the negative values of $T_{J_z}$ in the late stage of the
evolution, indicating the presence of a left-moving component, most
likely due to a spurious reflection at the outlet boundary.

In Fig.~\ref{fig:imp05_v50_m1}, we show selected snapshots from the
first $1800$ time steps of the simulation for impurity percentage
$C=0.5\%$ and $V=50$ meV. From this figure, we observe that a portion
of the wave packet gets ``trapped'', moving at lower speed, within the
impurity medium, while another portion manages to move out faster.
 
\subsection{Momentum Transmission Coefficient $T_{Jz}$}

\begin{figure}
  \centering   
  \includegraphics[scale=0.35]{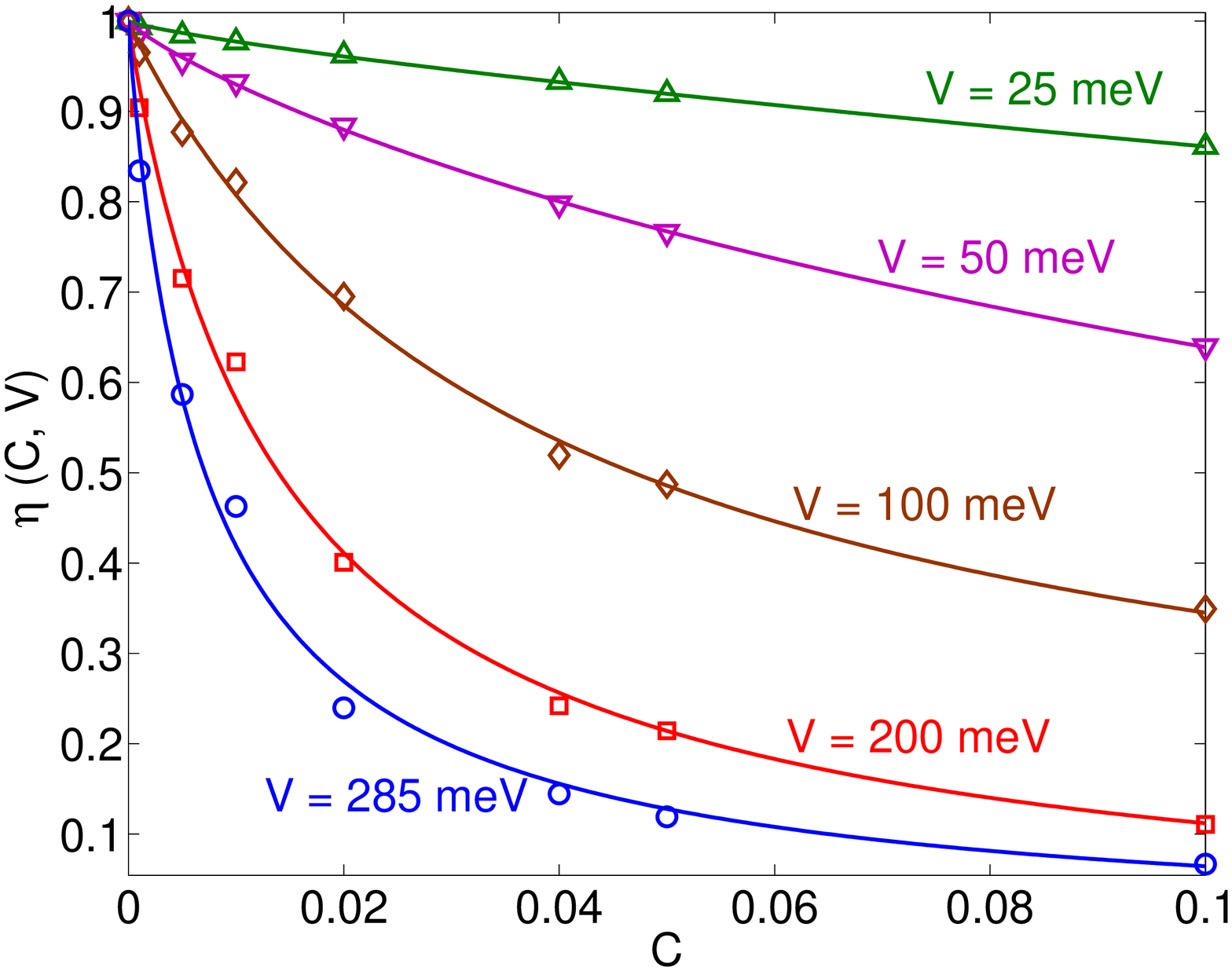} \\
  \includegraphics[scale=0.35]{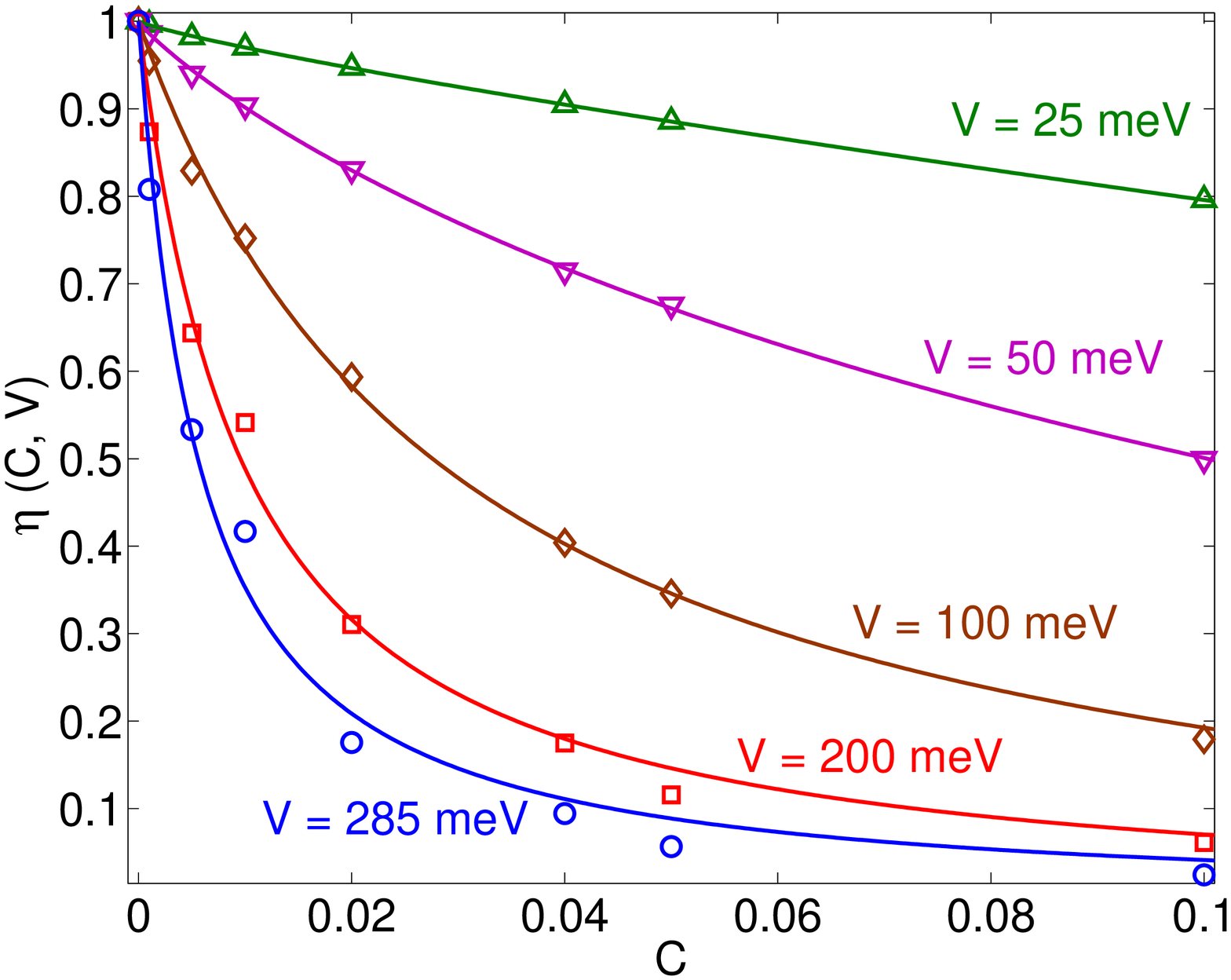} \\
  \includegraphics[scale=0.35]{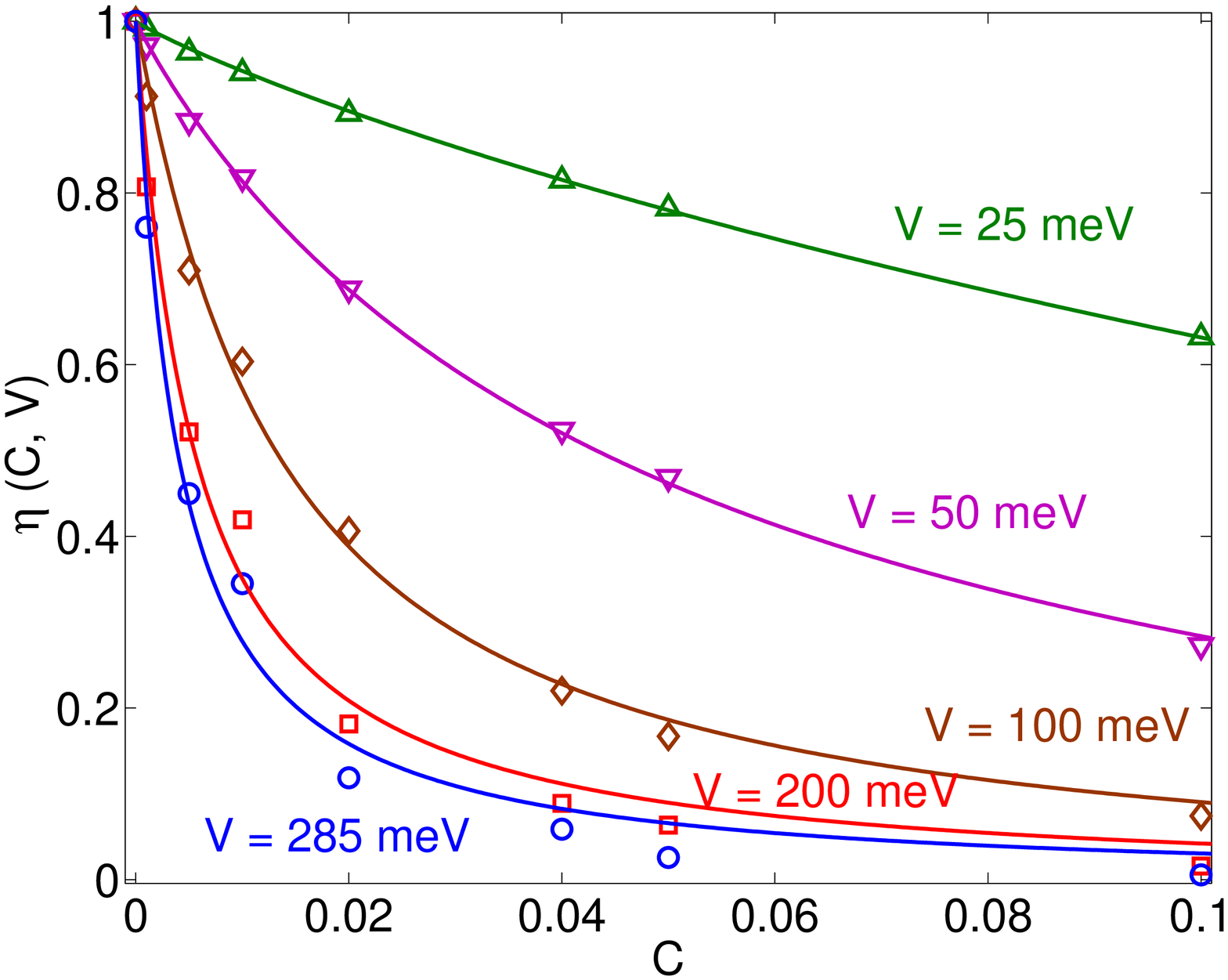} 
  \caption{Maximum value of $T_{Jz}$ as a function of the impurity
    percentage for each value of the impurity potential $V=50 \div
    285$.  For three values of the mass, $m=0$ (top), $0.05$ (middle),
    $0.1$ (bottom).}
  \label{fig:max_Jz}
\end{figure} 

In order to summarize the results obtained in the previous sections,
we inspect the maximum of the transmission coefficient $T_{Jz}$ in
Figs.~\ref{fig:m0_jz_v} and \ref{fig:m1_jz_v}, as a function of the
impurity potential and concentration, for three different values of
mass, $m=0, 0.05, 0.1$ (see Fig.~\ref{fig:max_Jz}).  These data
summarize the loss of momentum, hence resistivity, due to the random
impurities, formally measured by
\begin{equation}
\label{CONDU}
\eta(C,V) = \max(T_{J_z}(C,V)) \quad .
\end{equation}
From these figures, we observe that at high impurity concentration,
$C=0.05$, and a barrier $V=100$ meV, the relativistic wavepacket
looses about $50\%$ of its momentum, as compared the case of a pure
sample ($C=0$).  At the same concentration, a massive wave packet with
$m=0.1$, would loose more than $80 \%$, indicating a significant drop
of transmissivity due to inertia. At low impurity level, $C=0.001$,
both massless and massive wave-packets show a mild reduction of
transmittivity, below $10 \%$.

Let us now define the following ``transmittance'':
\begin{equation}
\label{CONDU1}
\Sigma(C,V) \equiv \frac{\eta}{1 - \eta} \quad .
\end{equation}  
This definition allows to draw a quantitative parallel with the
concept of permeability of a classical fluid moving through a porous
medium.  That is, when the transmittance is unity, the conductivity
goes formally to infinity, whereas zero transmittance connotes zero conductivity.

Using Eq.~\eqref{CONDU1}, we have found that the numerical results
are satisfactorily fitted by the following analytical expression:
\begin{equation}
\label{CONDU1}
\Sigma(C,V) = A\frac{(1-C)^{n+1}}{C^n} + \Sigma_0 \quad ,
\end{equation}  
where $A, n, \Sigma_0$ are fitting parameters, which depend on the
strength of the potential and the mass of the particles.  In
Fig.~\ref{fig:max_Jz}, we report the results of the fitting (solid
line), showing good agreement with the numerical data. We have plotted
$\eta$ instead of $\Sigma$, in order to avoid the divergence at $C=0$.
The values of the parameters can be found in Table~\ref{table1}.
\begin{table}
  \centering
  \begin{tabular}{|c|c|c|c|c|c|c|}\hline
    & V (meV) & 25 & 50 & 100 & 200 & 285\\ \hline
    & A & 1.09 & 0.26 & 0.046 & 0.017 & 0.0097  \\ 
    m=0 & n & 0.8 & 0.85 & 0.98 & 0.97 & 0.94 \\ 
    & $\Sigma_0$ & 0.51 & 0.23 & 0.17 & 0 & 0 \\ \hline
    & A & 0.68 & 0.16 & 0.03 & 0.009 & 0.005  \\ 
    m=0.05 & n & 0.84 & 0.88 & 0.99 & 1.01 & 1.01 \\ 
    & $\Sigma_0$ & 0 & 0 & 0 & 0 & 0 \\ \hline
    & A & 0.27 & 0.053 & 0.011 & 0.0053 & 0.0039  \\ 
    m=0.1 & n & 0.89 & 0.96 & 1.04 & 1.01 & 1.00 \\ 
    & $\Sigma_0$ & 0 & 0 & 0 & 0 & 0 \\ \hline
 \end{tabular}
 \caption{Set of parameters that has been obtained by fitting the numerical results for
   $\Sigma$ using Eq.\eqref{CONDU1}.}
  \label{table1}  
\end{table}
From this Table, we appreciate that the residual $\Sigma_0$, is zero
when the mass is different from zero, which points to this minimum
permeability (conductivity) as to a property of massless particles.
On the other hand, massive particles show a closer adherence to the
Kozeny-Carman law, in the context of classical fluid dynamics
\cite{rumer, Bear}, where no residual conductivity is observed at
$C=1$.  Also, note that for low potential barriers, the exponent is
around $n \sim 0.85$, while for intermediate and strong potentials it
is near $n \sim 1$, i.e. the value it takes for classical fluid
dynamics in a dilute disordered medium.  Thus, for strong potentials,
the classical analogy shows satisfactory results, while for intermediate
and weak potentials, it presents deviations, typically of the order of $15\%$.
Finally, we observe that the case $m=0$ shows a significantly higher
transmission than the corresponding data with $m>0$, which is due to
the higher momentum losses in the impurity region.
It appears plausible to interpret the non-negligible surplus of relativistic conductivity, especially 
for the three cases with Fermi energy $E_F<V$, as an indirect manifestation of Klein tunneling.

\section{Conclusions and Discussion}

In this paper we have performed a numerical study of a relativistic
Gaussian wave-packet propagating through a disordered medium, which we
modeled as a set of randomly located potential barriers.

From the numerical results, we conclude that for high concentration of
impurities, the wave packet presents higher losses in momentum.
Furthermore, for a given impurity concentration, by increasing the
potential of each impurity, we also find a loss of momentum.  Systems
with massive excitations are also studied, which can be of relevance
to the case of gaped graphene samples.  A non-zero mass is found to
produce higher losses of momentum in the impurity region.  The actual
numerical values show that at high impurity concentration, $C=0.05$,
the wavepacket looses more than half of its momentum with barriers of
$100$ meV and up to $85 \%$ with $V=285$ meV.  At low concentrations,
$C=0.001$, however, the losses are much milder, going from about
$5-20\%$, for $V=100$ to $285$ meV, respectively.

These data can be regrouped into an analytical expression, which bears
a strong similarity with the permeability of porous media, as a function of the
porosity. We have estimated the value of the conductivity from the
transmission coefficient and fitted it by using the Carman-Kozeny law
for porous media, relating the permeability with the concentration of
impurities We have found that this analogy works pretty well for the
massive case, which shows no residual conductivity and a scaling
exponent pretty close to unity. On the other hand, the massless case
shows a residual conductivity, which can possibly be related to the minimum
conductivity of graphene. Moreover, for weak and intermediate
potential strengths, the exponent is not unity, corresponding to a
fractional Kozeny-Carman law.  On the other hand, for strong
potentials, the exponent $1$ is recovered to a good accuracy, bringing
the results closer to the analogy with classical fluids \cite{DiVentra}.  
The applicability of this classical analogy indicates that, at least for
the parameter set investigated in this paper, quantum tunneling is not
the dominant transport mechanism, as compared to the semi-classical
dynamics of the wave-function, which can turn around the obstacles in
a similar way as a classical fluid would do.
The results of this paper are expected to be amenable to experimental
validation. For this purpose, samples of graphene with local chemical
doping could be used \cite{NATURE, doping}. In addition, for
validating the results with massive particles, a substrate of SiC will
also be required, in order to generate the gap due to the presence of
particle mass. Finally, as a byproduct, we have introduced a new tool
to model electronic transport in graphene, namely the quantum lattice
Boltzmann method (QLB). QLB shares a remarkable computational
efficiency, especially on parallel computers, and easy handling of
complex geometries with its well-established classical LB
counterpart. As a result, it is hoped and expected that the present
model can make a contribution to the computational study of transport
phenomena in graphene and other physical systems governed by the Dirac equation.

\begin{acknowledgments}
  The authors are grateful for the financial support of the
  Eidgen\"ossische Technische Hochschule Z\"urich (ETHZ) under Grant
  No. 06 11-1.
\end{acknowledgments}

\end{document}